**Title**

Critical assessment of contact resistance and mobility in tin perovskite semiconductors

**Author list**

Youcheng Zhang[1], Stefano Pecorario[1], Xian Wei Chua[1,2], Xinglong Ren[1], Cong Zhao[1], Rozana Mazlumian[1], Satyaprasad P. Senanayak[3], Krishanu Dey[1,2], Sam Stranks[1,2] and Henning Sirringhaus[1]*

[1]*Optoelectronics Group, Department of Physics, Cavendish Laboratory, University of Cambridge, Cambridge, UK*
[2]*Department of Chemical Engineering and Biotechnology, University of Cambridge, Cambridge, UK*
[3]*Nanoelectronics and Device Physics Lab, National Institute of Science Education and Research, School of Physical Sciences, HBNI, Jatni, India*

**Abstract**

Recent reports highlight the potential of tin-based perovskite semiconductors for high-performance *p*-type field-effect transistors (FETs) with mobilities exceeding 20 cm² V⁻¹ s⁻¹. However, these high mobilities—often obtained via two-probe (2P) methods on devices with small channel length-to-width ratios (*L/W* < 0.5) operating in the saturation regime at high drain-source currents—raise concerns about overestimation due to contact resistance and non-ideal FET characteristics. Here, we performed gated four-point probe (4PP) FET measurements on Hall bar devices (*L/W* = 5) of $\mathrm{Cs_{0.15}FA_{0.85}SnI_3}$, obtaining a consistent mobility of 3.3 cm² V⁻¹ s⁻¹. Upon comparing these with gated 2P measurements of narrow-channel FETs (*L/W* = 0.1) on the same chip, we resolved the contact resistance ($R_C$). The 2P linear mobility is underestimated due to voltage drops across $R_C$, while the 2P saturation mobility is overestimated because of high $(\frac{\partial R_C}{\partial V_G})$ near the threshold. Contact resistance effects become more pronounced at lower temperatures. Contact-corrected 4-point-probe (4PP) mobilities are independent of bias conditions and are observed to flatten at temperatures lower than 180 K. Future reports of perovskite FET mobilities should include gated 4PP measurements and use devices with larger *L/W* ratios to minimize nonidealities arising from contact resistance effects.

# Introduction

The performance of perovskite field-effect transistors (FETs) has seen significant improvements in recent years. Recent studies highlight the exceptional potential of tin-based perovskites for high-performance p-type transistors, with field-effect mobility values exceeding 20 cm$^2$V$^{-1}$s$^{-1}$.[1] These impactful findings are expected to inspire further research efforts aimed at broader applications for tin perovskite-based FETs.

The fabrication of perovskite FETs is often based on methods that are widely used for organic FETs (OFETs), typically involving solution-processed deposition of perovskite thin films and thermal evaporation of metal contacts. Source and drain electrodes form a channel geometry with channel length (*L*) of 20 to 200 µm and channel width (*W*) around 1 mm. Transfer and output characteristics are measured by applying a source-drain bias ($V_{DS}$) and a gate bias ($V_G$). This method is referred to as gated Two-Probe (2P) method, where apart from the gate, only source and drain are connected to the device. FET mobility values are extracted from transfer characteristics in the saturation or linear regime. The Two-Probe method has been commonly adopted for recent high mobility reports on tin perovskite FETs.[1–8]

In the 2P method, any charge carrier injection barrier at the source-drain metal-semiconductor interface manifests itself as a contact resistance ($R_C$) in the measured electrical characteristics. Consequently, the source-drain current $I_{DS}$ which ideally follows $I_{DS} = V_{DS}/R_S$ (where $R_S$ is the sheet resistance of the semiconductor channel) is now modified to $I_{DS} = V_{DS}/(R_S + R_C)$ to account for the contribution of $R_C$, if we assume Ohm's law to be valid. Thus, commonly used 2P method cannot directly resolve contact resistance.

In the linear and saturation regimes, mobilities are calculated by the following formulae, where $C_i$ is the internal capacitance of the dielectric layer:

$$\mu_{FET,lin} = \frac{L}{W \cdot C_i} \cdot \frac{1}{V_{DS}} \cdot \frac{\partial I_{DS}}{\partial V_G}$$

*(1)*

$$\mu_{FET,sat} = \frac{2L}{W \cdot C_i} \cdot \left(\frac{\partial \sqrt{I_{DS}}}{\partial V_G}\right)^2$$

*(2)*

Term $\left(\frac{\partial I_{DS}}{\partial V_G}\right)$ and $\mu_{FET,lin}$ can be derived as a function of $R_C$ and $R_S$:

$$\frac{\partial I_{DS}}{\partial V_G} = V_{DS} \cdot \frac{\partial(1/(R_C + R_S))}{\partial V_G} = -\frac{V_{DS}}{(R_C + R_S)^2} \cdot \left(\frac{\partial R_S}{\partial V_G} + \frac{\partial R_C}{\partial V_G}\right)$$

*(3)*

$$\mu_{FET,lin} = \frac{L}{W \cdot C_i} \cdot \frac{1}{(R_C + R_S)^2} \cdot \left(\frac{\partial R_S}{\partial V_G} + \frac{\partial R_C}{\partial V_G}\right)$$

*(4)*

Gundlach et al. reported that $R_C$ can be modulated by gate bias ($V_G$). A significant contribution from the term $\frac{\partial R_C}{\partial V_G}$ in the threshold region can lead to an overestimation of $\mu_{FET,lin}$ and $\mu_{FET,sat}$ by a large factor, that can exceed 3[9] However, if $R_C$ is relatively independent of $V_G$, the term $\frac{\partial R_C}{\partial V_G}$ could be neglected. In this case, if $R_C$ is large, it will also result in a small $\frac{\partial I_{DS}}{\partial V_G}$, as suggested by the coefficient $\frac{V_{DS}}{(R_C+R_S)^2}$. Therefore, it is essential to carefully consider the contributions of both $R_C$ itself and $\frac{\partial R_C}{\partial V_G}$ when extracting FET mobility using the commonly employed 2P method.

For tin perovskite FETs, Podzorov et al. recently emphasized the importance of reporting transfer curves measured in the linear regime at low $I_{DS}$, as operating in the saturation regime with high channel current often leads to uncontrolled nonlinear device characteristics.[10] Several recent reports on tin perovskites showing FET mobilities greater than 20 cm$^2$V$^{-1}$s$^{-1}$ are based on 2P method with a small channel length-to-width ratio (L/W < 0.5). These mobility values are extracted from the saturation regime under electric field strength exceeding 200 V/mm and $I_{DS}$ close to 10 mA. This has sparked debates regarding the practicability of such extreme biasing condition and whether the high mobility values may be overestimated.[11,12]

Hence, gated four-point probe (4PP) measurements are commonly utilized to extract the mobility and eliminate the influence of contact resistance. The layout of a 4PP device is shown in Figure S1 where two additional narrow electrodes are used to probe the voltage drop along the channel. The semiconductor layer is patterned into a rectangular bar (referred to as a Hall bar structure) to prevent electrical pathways through the contact electrodes. The 4PP transfer characteristics are measured while simultaneously recording voltage values from the two voltage probes. The device is operated in the linear regime to ensure a uniform voltage drop along the channel and to avoid channel pinch-off. 4PP FET mobility ($\mu_{FET,4PP}$) can be calculated by formula (5), where $D_{4P}$ and $V_{4P}$ are the distance and voltage difference between the two voltage probes.

$$\mu_{FET,4PP} = \frac{D_{4P}}{W \cdot C_i} \cdot g_{m,4PP} = \frac{D_{4P}}{W \cdot C_i} \cdot \frac{\partial(I_{DS}/V_{4P})}{\partial V_G}$$

*(5)*

By replacing $V_{DS}$ with the voltage drop across the two voltage probes in the channel ($V_{4P}$), the contributions from $R_C$ and term $\frac{\partial R_C}{\partial V_G}$ can be eliminated. This allows for the extraction of linear regime FET mobility without the influence of contact resistance and within a safe current range. Therefore, debates regarding the overestimation of FET mobility in recent tin perovskite reports could be resolved by performing gated-4PP measurements on Hall bar geometry FETs with a large channel length-to-width ratio (L/W > 2). However, the sensitivity of perovskites to water and oxygen has made it challenging to pattern solution-processed perovskite films using common photoresist methods, thus limiting the range of charge transport measurements that can be performed on these materials, including gated-4PP FET measurements.

In this study, we performed gated four-point-probe (4PP) FET measurements on patterned Hall bar devices based on tin perovskite with the composition $Cs_{0.15}FA_{0.85}SnI_3$. The devices demonstrate a 4PP mobility of 3.3 cm$^2$V$^{-1}$s$^{-1}$ at room temperature, which remains independent of the source-drain bias ($V_{DS}$) but shows a weak dependence on the gate voltage ($V_G$). We found that the gate bias-dependency arises from ion migration, which can be mitigated by employing a pulsed gate bias or conducting measurements at low temperatures. Additionally, on the same chip, gated two-probe (2P) measurements of narrow channel FETs (*L/W* = 0.1) exhibit comparable mobility values but show a more significant dependence on $V_{DS}$ and $V_G$. Through the extraction of contact resistance and channel sheet resistance from the 4PP measurements, we found these dependences originate from contact resistance. We also found that contact resistance is asymmetric such that the source side resistance is much larger than the drain side resistance. Compared to the 4PP results, the 2P linear mobility measured with low $V_{DS}$ is underestimated due to a substantial voltage drop across $R_C$, while the 2P saturation mobility measured at high $V_{DS}$ are overestimated from the term $\frac{\partial R_C}{\partial V_G}$. This discrepancy also leads to differences in the temperature dependence of the 4PP and 2P mobilities in both linear and saturation regimes. Specifically, 2P linear mobilities are more significantly underestimated at low temperatures. In contrast, the 4PP mobility exhibits a consistent temperature dependence trend governed by thermal activation, with activation energy of 75$\pm$10 meV and remains independent of $V_{DS}$. 4PP mobility flattens and slightly increases at temperatures below 180K, which might suggest band-like conduction governed by impurity scattering when thermal energy is not sufficient to drive thermal activation transport.

Our work demonstrates the effectiveness of gated-4PP FET measurements on tin perovskites using a Hall bar geometry, enabling more accurate extraction of the mobility as a function of temperature compared to the conventional gated Two-Probe method. This not only provides a more robust assessment of transistor performance, but also provides more reliable insight into the underpinning charge transport physics of tin perovskite semiconductors.

## Results

### Experiments and Materials

We select tin perovskite composition $Cs_{0.15}FA_{0.85}SnI_3$ (with 12.5mol% extra SnF$_2$) for this study because a $Cs_{0.15}FA_{0.85}$ ratio exhibited the highest mobility and hysteresis-free transfer characteristics in our previous work on mixed Pb-Sn perovskites.[13] We first characterised the material properties of spin-coated $Cs_{0.15}FA_{0.85}SnI_3$ films. In Figure 1a, the as-deposited perovskite film exhibits a smooth surface morphology with grain size of approximately 1 μm. The X-ray diffraction (XRD) pattern in Figure 1b confirms a pure γ-phase orthorhombic structure by distinct (101) and (202) peaks.[14] In Figure 1c, both UV-Vis absorption and photoluminescence (PL) emission spectra indicate a band gap of 1.34 eV. Time-resolved PL measurements reveal a *1/e* lifetime of 2.9 ns at low excitation fluences (Figure 1d). These material and photophysical properties are consistent with previous photophysical reports on $Cs_{0.15}FA_{0.85}SnI_3$.[15–17]

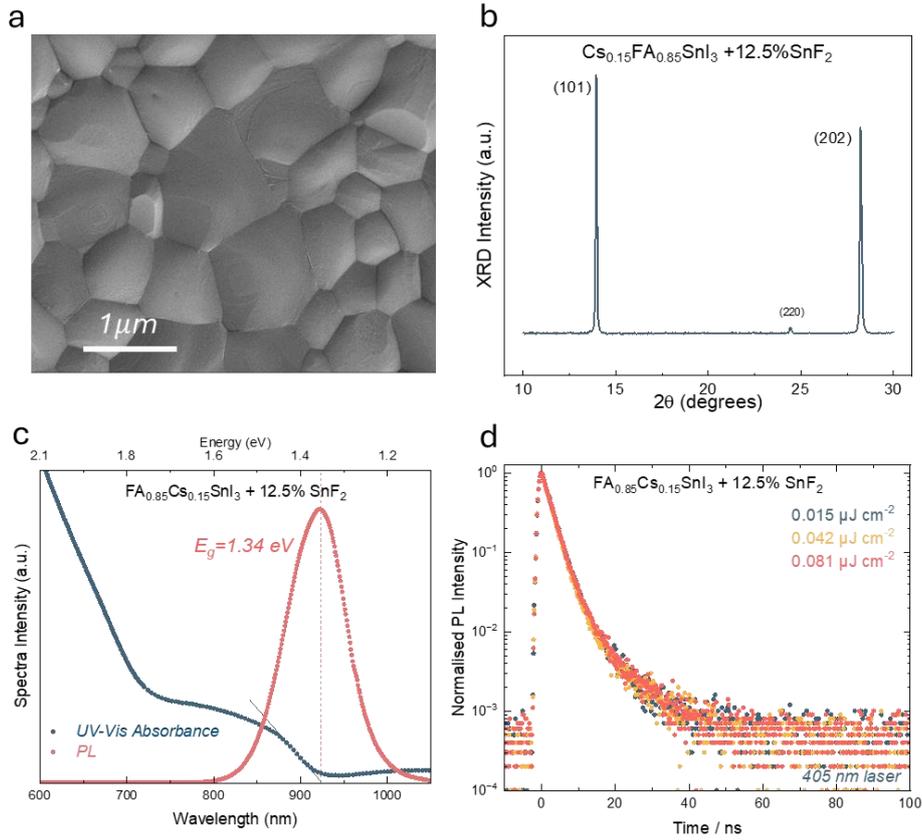

*Figure 1 Material characterisations for $Cs_{0.15}FA_{0.85}SnI_3$. **(a)** Scanning electron microscopy (SEM) image, **(b)** $2\theta$ XRD pattern, **(c)** UV-Vis absorption and photoluminescence (PL) emission spectra and **(d)** PL decay kinetics obtained by time correlated single photon counting at low fluences.*

**Gated Four-Point Probe FET Characteristics**

For the fabrication of FET devices, we used a bottom-gate, bottom-contact architecture: Perovskite films were spin-coated on a pre-patterned SiO₂/Si substrate with metal contacts of 2P and 4PP FETs and then covered by an encapsulation layer of PMMA, that was used to protect the film against accidental air/oxygen exposure. The film was first rough-patterned by cocktail sticks in a glovebox and then fine-patterned by mechanical scratching the films using metallic tips in a vacuum probe station. Mechanical patterning ensured the absence of any chemical degradation of the sensitive $Cs_{0.15}FA_{0.85}SnI_3$ films. The device chip containing both 2P and 4PP devices is shown in Figure 2a, with the top left showing an overview of the device chip and the other panels showing individual 2P and 4PP FET devices after patterning. The 4PP FET device consists of a Hall bar patterned channel with a channel length of 600 μm and a channel width of 120 μm and eight voltage probes (Figure 2a top-right panel or Figure S1). The voltage probes closest to the source and drain electrodes are used in the measurement of $V_{4P}$ with $D_{4P} = 570$ μm, marked with $V_{P1}$ and $V_{P2}$ in Figure 2a top-right panel.

FET characteristics of the 4PP device are shown in Figure 2b-g. The output curves are hysteresis free and exhibit good linearity (Figure 2b), indicating that the device operates in the linear regime at -20 V<$V_{DS}$<0 V. Figure 2c presents the $I_{DS}$ plots of the transfer characteristics. The curves show ON/OFF ratios of $10^3$ and turn-on voltages ($V_{ON}$) exceeding +80 V. When $V_{DS}$ increases to -18 V, turn on voltage increases to nearly +100 V. The high-turn on voltages are typical for these tin-based perovskites and reflect unavoidable p-type doping of the films arising from tin vacancy defects.[3,14,18] The vacancy defects ($V_{Sn}$) trap electrons and form $V_{Sn}^-$ or $V_{Sn}^{2-}$, thereby increasing the number of free holes in the material.[19] Tin vacancies can easily form due to their low formation energy,[14] and the oxidation of $Sn^{2+}$ to tin oxide products also promotes the formation of tin vacancies.[19,20]

Notably, near $V_G$ = 0 V the hysteresis of the transfer characteristics shifts from counterclockwise (CCW, for positive $V_G$) to clockwise (CW, for negative $V_G$). This unusual behaviour is also clear in the $I_{DS}$ linear plots of the transfer curves shown in Figure S2. CCW hysteresis may arise from the screening of gate voltage by ion migration or trapping of charge carriers at the dielectric interface. The origin of the CW hysteresis is unclear. As discussed below, it seems to be related to the observed increase in mobility with increasing gate voltage, which is also a consequence of ion migration, as it disappears when measuring in pulsed mode or at low temperature.

Voltage values measured from the two voltage probes ($V_{P1}$ and $V_{P2}$) normalized by $V_{DS}$ are shown in Figure 2d. $V_{P2}$ closer to the source electrode displays a larger voltage drop and stronger modulation by both $V_G$ and $V_{DS}$, while $V_{P1}$ is nearly equal to $V_{DS}$. This indicates that contact resistances are asymmetrical at source and drain sides; the source side resistance is much larger than the drain side resistance near the threshold region. This may suggest that in the ON state the contact resistance is dominated by the reverse biased Schottky diode at the source junction, while the forward-biased metal-semiconductor junction at the drain side is less resistive. [21–24]

4PP transfer characteristics plot in conductance (Figure 2e) demonstrate a good overlap of conductance for curves measured at different $V_{DS}$. The transfer curves are plotted on a linear scale in Figure 2f, displaying a slightly superlinear increase in channel conductance with rising $V_{DS}$. This nonideality may be attributed to channel joule heating effect or to the same underlying ion migration induced *p*-type doping mechanism responsible for the CW hysteresis.

Figure 2g shows the calculated 4PP FET linear mobility from the forward transfer scan, ranging from 2 $cm^2V^{-1}s^{-1}$ at positive $V_G$ to 5 $cm^2V^{-1}s^{-1}$ at negative $V_G$. Despite this dependence on $V_G$, mobility values obtained at different $V_{DS}$ consistently overlap, indicating that the results are unaffected by contact resistance. Overall, the mobility exhibits an average value of 3.3±1.2 $cm^2V^{-1}s^{-1}$. This is the mobility value averaged from data points between -40V<$V_G$<80V in Figure 2g.

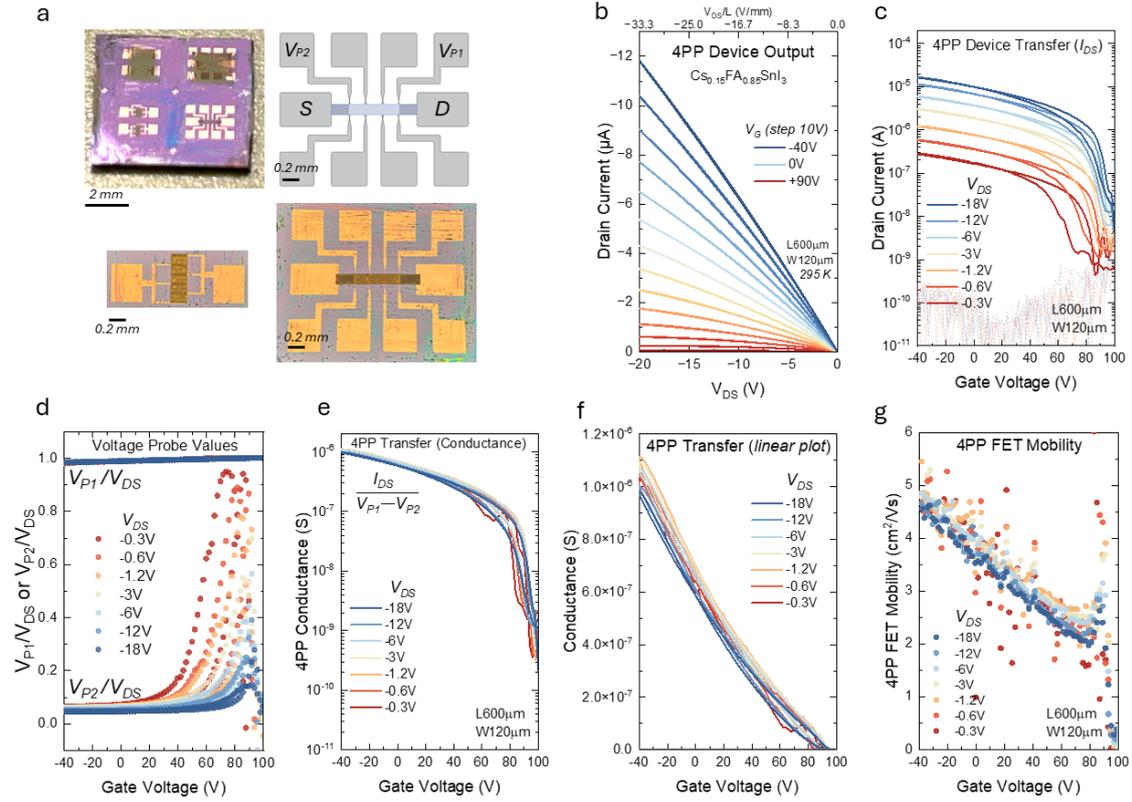

*Figure 2 Gated four-point-probe (4PP) FET Devices and Characteristics.*

**(a)** Device chip containing both 2P and 4PP FET devices. Top-left shows the SiO$_2$ chip with two 2P and one 4PP device patterned at the bottom. Top-right shows the geometry of the 4PP device with electrodes marked in grey and channel Hall-bar marked in dark blue. Source (*S*), drain (*D*), voltage probes ($V_{P1}$ and $V_{P2}$) are marked on the electrodes. Bottom-left and right shows the optical microscopy image of the patterned 2P and 4PP FET devices for measurement. **(b)** Output characteristics measured at different $V_G$. **(c)** 4PP device $I_{DS}$ plots of transfer characteristics measured at different $V_{DS}$. Gate leakage current at different scans is shown in faint dash line. **(d)** Gate modulation of voltage probe values ($V_{P1}$ and $V_{P2}$) normalized by $V/V_{DS}$. Raw data of the voltage probe values are plotted in Figure S3. **(e)** 4PP transfer characteristics in conductance. Conductance is calculated by $\frac{I_{DS}}{V_{P1}-V_{P2}}$. **(f)** Transfer characteristics (conductance) plotted in linear scale. **(g)** Extracted 4PP mobility as a function of gate voltage from forward scan transfer curves. *L*= 600 μm, *W*=120 μm, $D_{4P} = 570$ μm. Transfer characteristics were measured with continuous gate bias at 295K. $V_{DS}$ is controlled within -20 V to avoid channel pinch-off and SiO$_2$ dielectric breakdown (due to the large contact pad area).

## Suppression of $V_G$-dependence by pulsed gate bias and low temperature

The $V_{DS}$-independent mobility indicates that the 4PP measurement effectively avoids the influence of contact resistance. However, a textbook-like FET device should exhibit mobility independent of both $V_G$ and $V_{DS}$. The 4PP method cannot eliminate the non-idealities arising from dielectric interface. Reports on lead-based perovskite FETs found that ion migration could be suppressed at low temperatures.[25,26] Additionally, a pulsed gate bias could reduce ion migration in the vertical direction at the dielectric interface.[27,28] These methodologies could also mitigate ion migration in tin perovskite devices. Therefore, we applied both approaches to see whether it is possible to improve the linearity of the transfer curve and extract mobility values less dependent on $V_G$.

Figure 3a,b,c present the transfer characteristics and mobility measured using a pulsed gate at room temperature. The transfer curves appear more linear and display an early turn-off near +80 V. The mobility stabilizes at 2.6 cm$^2$V$^{-1}$s$^{-1}$ for $V_G$< +20 V, which is slightly lower than the values obtained from continuous mode measurements. The $V_G$-independent mobility indicates that pulsed gate bias can effectively eliminate the nonidealities exhibited in continuous mode measurements. FET measurements were also conducted at 220 K using a continuous gate bias. The transfer curves show excellent linearity with minimal hysteresis (Figure 3d,e). Mobility also stabilizes at 1.1 cm$^2$V$^{-1}$s$^{-1}$ for different $V_G$ and $V_{DS}$ values in Figure 3f, indicating that ion migration is effectively supressed at 220 K. This observation further confirms that the $V_G$-dependence seen at room temperature can most likely be attributed to ion migration effects. It is less likely to arise from joule heating, as joule heating effect would be more pronounced at low temperatures.

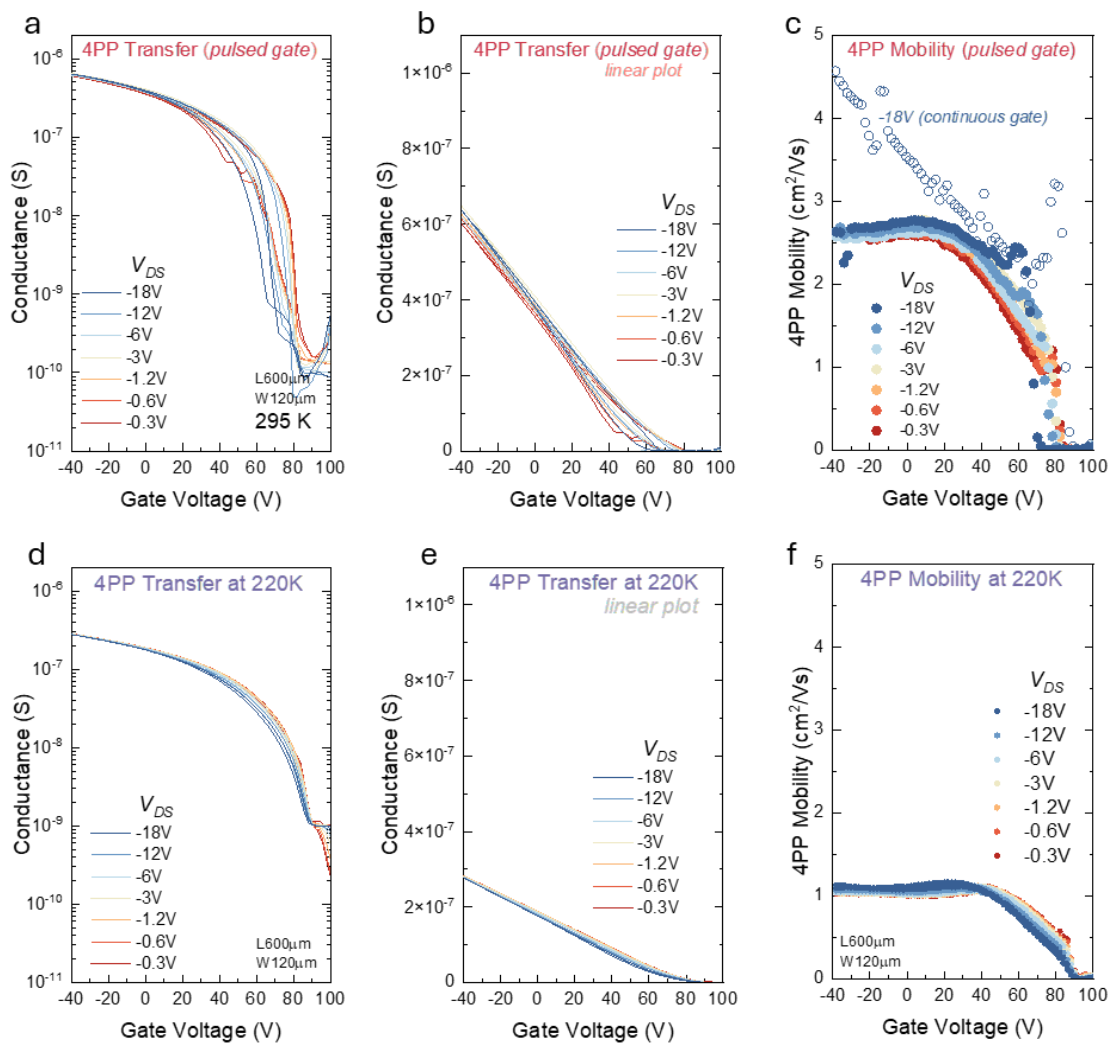

*Figure 3 4PP FET Characteristics with Mitigated Impact of Ion Migration*

**(a)** Transfer characteristics measured with pulsed gate bias at 295 K. **(b)** Transfer curves plotted on a linear scale. **(c)** Extracted 4PP mobility from pulsed gate measurement. Hollow circles represent 4PP mobility extracted from measurement with continuous gate bias. **(d)** Transfer characteristics measured at 220 K with continuous gate bias. **(e)** Transfer curves plotted on a linear scale. **(f)** Extracted 4PP mobility at 220 K (forward and reverse scans). Device is the same as in Figure 2. $I_{DS}$ plots of the transfer curves are shown in Figure S2.

**Gated two-probe FET characteristics**

A two-probe (2P) FET device with a channel length of 100 μm and channel width of 1 mm was also measured on the same chip in continuous gate-bias. Figure 4 presents the FET characteristics of the 2P device. In the linear regime, the transfer characteristics show a turn-on voltage at +80 V, an ON-OFF ratio ranging from $10^5$ to $10^7$, and a moderate level of hysteresis (Figure 4a). The transfer curves are slightly non-linear and exhibit a change in hysteresis loop direction from CCW to CW (Figure 4b). In the saturation regime, the transfer characteristics display a high ON current exceeding 2 mA and an ON/OFF ratio of $10^7$ (Figure 4c), with turn-on voltages greater than +90 V. 2P output characteristics clearly show the transition from saturation to the linear regime as gate voltage changes from positive to negative (Figure 4d). The extracted linear regime 2P FET mobility in Figure 4e displays a strong dependence on both $V_G$ and $V_{DS}$, ranging from 2 to 4 $cm^2V^{-1}s^{-1}$ at $V_G$ =-40 V. The dependence is even more pronounced in the saturation regime, as shown in Figure 4f. Mobility scales with $V_{DS}$, reaching a peak of 5.5 $cm^2V^{-1}s^{-1}$ around $V_G$ = +45 V for $V_{DS}$= -60 V, before sharply dropping to 3.5 $cm^2V^{-1}s^{-1}$ at $V_G$ = 0 V. Figure S4 demonstrates that cooling to 220 K reduces hysteresis in the transfer characteristics. However, the device becomes more susceptible to contact resistance effects, as indicated by the mobility hump near the threshold region and the superlinear line shape of the output curves.

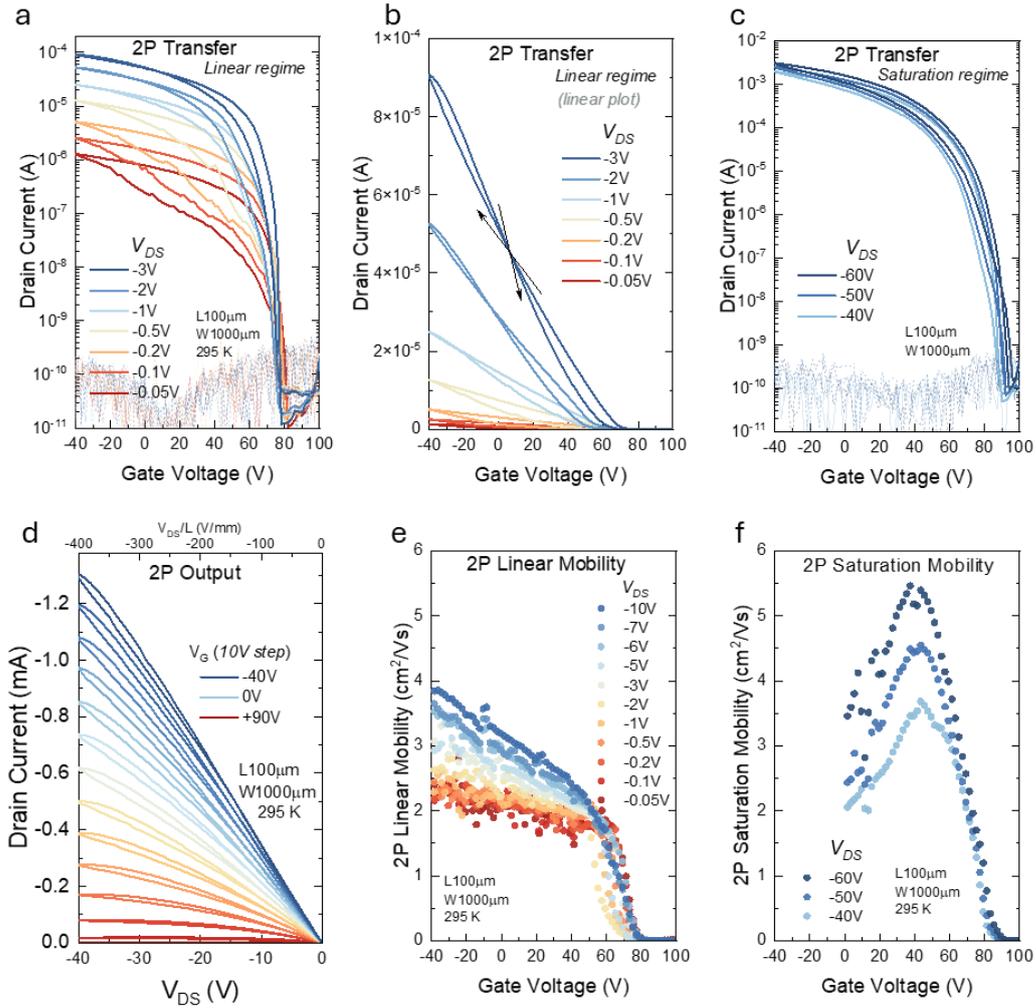

*Figure 4 2P FET characteristics and mobility*

**(a)** Linear regime transfer characteristics and **(b)** plots on a linear scale. Hysteresis loop direction is marked by arrows. **(c)** Saturation regime transfer characteristics. **(d)** Output characteristics. Extracted 2P linear regime mobility **(e)** and saturation regime mobility **(f)** as a function of $V_G$ measured at different $V_{DS}$ values for the forward scans. Saturation regime mobility is only plotted for $V_G > 0$ V due to a transition to linear regime at $V_G < V_{ON} + V_{DS}$, where $V_{ON}$ is the turn-on voltage from transfer curve. FET characteristics were measured using continuous gate bias. 2P device with $L$=100 µm and $W$=1 mm is on the same chip as the 4PP device shown in Figure 2.

## 2P Mobility vs. 4PP Mobility

These results suggest that while the device performance of the 2P device here is comparable to state-of-the-art tin perovskite FETs[13], its strong dependence on gate and source-drain voltages could lead to inaccuracies in reporting mobility metrics. In contrast, the 4PP devices show more consistent results with minimal dependence on source-drain voltage. Figure 5 compares the mobilities of 4PP and 2P devices against source-drain bias normalised by channel length ($V_{DS}/L$). Despite the large error bars in 4PP mobility due to signal noise and ion migration induced $V_G$-dependence, 4PP mobility remains consistent at around 3.3 $cm^2V^{-1}s^{-1}$ over an electric field strength range of 0.5 V/mm to 30 V/mm. In comparison, 2P linear mobility remains stable at approximately 2 $cm^2V^{-1}s^{-1}$ for electric field strength below 10 V/mm but increases rapidly to 3.5 $cm^2V^{-1}s^{-1}$ at 200 V/mm. 2P linear mobility values are extracted from the mobility versus gate voltage curves shown in Figure 4 for data points between $V_G$= -40V and +40V.

The average 2P saturation mobility values are extracted by fitting the $\sqrt{I_{DS}}$ versus $V_G$ curves shown in Figure S5. They exhibit strong modulation by electric field strength from 3 to 4.7 $cm^2V^{-1}s^{-1}$. Note that $V_{DS}$ higher than -20 V (-33.3 V/mm) could not be applied to the 4PP devices due to channel pinch-off (Figure S6). These observations were further validated by a same set of measurements from another chip (Figure S7). Thus, the comparison suggests a potential underestimation of 2P linear regime mobility and an overestimation of 2P saturation regime mobility, both of which could be influenced by the contact resistance effect. Figure S5c displays the comparison of 2P and 4PP mobilities at 220K, also showing a lower 2P linear mobility. The error bars for both 2P and 4PP are much smaller, confirming ion migration effect in the vertical direction is suppressed at low temperature.

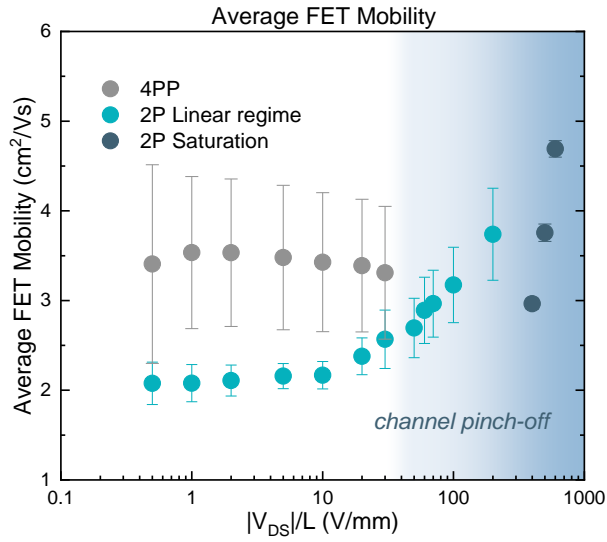

*Figure 5 Comparison of 2P and 4PP mobility extracted at different FET channel electric field strengths ( $|V_{DS}|/L$). 4PP mobility values are averaged at -40V<$V_G$<+80V for each $V_{DS}$. 2P linear regime mobility is averaged at -40V<$V_G$<+40V. 2P saturation regime mobility is obtained by fitting the $\sqrt{I_{DS}}$ versus $V_G$ curves. Error bars are standard deviation values of averaging or line fitting. Blue shade colour marks onset of channel pinch-off.*

**Contact Resistance Analysis**

Equation (3) suggests that the inconsistent 2P mobility could arise from contact resistance via term $\frac{V_{DS}}{(R_C+R_S)^2}$ and term $\frac{\partial R_C}{\partial V_G}$. 4PP output data can resolve $R_S$, $R_C$ and $R_{tot}$(=$R_C + R_S$) using the following equations:

$$R_S = \frac{\partial V_{4P}}{\partial I_{DS}} \cdot \frac{L}{D_{4P}}$$

*(6)*

$$R_C = \frac{\partial V_{DS}}{\partial I_{DS}} - \frac{\partial V_{4P}}{\partial I_{DS}} \cdot \frac{L}{D_{4P}}$$

*(7)*

$R \cdot W$ for $R_{tot}$, $R_S$ and $R_C$ extracted from the 4PP device output characteristics are plotted against $|\frac{V_{DS}}{L}|$ in Figure 6. Gate modulation reduces all three values as $V_G$ becomes negative. $R_S$ originates from both the channel dielectric interfacial region where the conductivity is modulated by gate bias, and a bulk region of the film away from the interface where the film's bulk conductivity dominates. A slight increase of $R_S$ is observed at $\frac{V_{DS}}{L}$>10 V/mm, indicating channel starts to pinch-off. $R_C$ is strongly modulated by $V_G$ and it could exceed $R_S$ in the OFF state. These effects contribute to the "U" shape behavior of $R_{tot}$ at $V_G$=+70 V and +60 V. Additionally, $R_C$ is also modulated by $V_{DS}$. Due to a reduction in the Schottky barrier width or enhanced thermionic emission, it decreases more rapidly also at $\frac{V_{DS}}{L}$>10 V/mm. A reduction of the contact resistance with more negative $V_G$ is commonly observed in p-type FETs and may reflect a more effective screening/lowering of the charge injection barrier when the accumulation layer is formed at the interface. However, in the present devices it is also important to consider that for positive $V_G$ the region of the semiconductor near the interface (dark mint) is depleted and charge transport is dominated by the remaining doped region (light mint) in the bulk of the perovskite film away from the interface (inset in Figure 6a). The depletion of the perovskite film near the interface could contribute to the increase in the contact resistance at positive $V_G$ as it might suppress charge injection from the side wall of the source-drain electrodes.

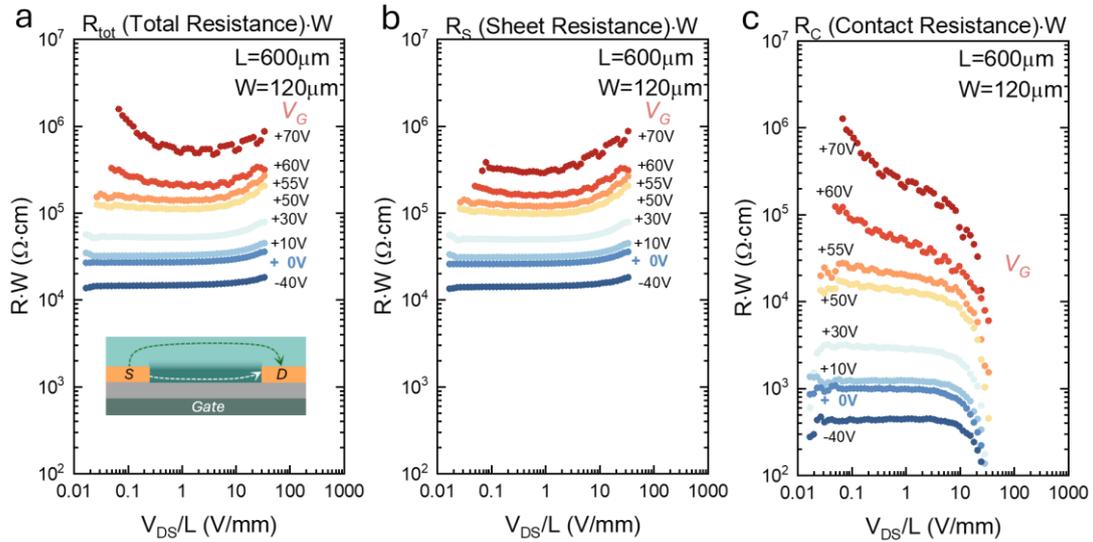

Figure 6 Extracted $R \cdot W$ for **(a)** total resistance $R_{tot}$, **(b)** channel sheet resistance $R_S$, and **(c)** contact resistance $R_C$ at different $V_G$ and $V_{DS}/L$ values. Resistance values are extracted from 4PP output characteristics using Equation (6) and (7). The inset in panel (a) illustrates the vertical geometry of a bottom-gate, bottom-contact FET. For positive $V_G$ the region of the semiconductor near the interface (dark mint) is depleted and charge transport is dominated by the remaining doped region (light mint) in the bulk of the perovskite film away from the interface. Channel length L is 600 μm and channel width W is 120 μm. 4PP distance $D_{4P}$ is 570μm. Output characteristics was not performed at $V_G$=+100V due to dielectric leakage. Raw data for output characteristics and change of $V_{4P}$ are shown in Figure S8.

Figure 7a,b,c shows the gate modulation of resistances at three typical channel field strengths (-0.11 V/mm, -1.1 V/mm, and -21 V/mm). It is clear that contact resistance is strongly modulated by gate voltage, particularly in the near-threshold region, and can eventually exceed $R_S$, suggesting injection barriers at the contact interfaces are extremely sensitive to charge carrier concentration modulated by gate bias. The fast decline in $R_C$ at more positive

gate voltages, compared to $R_S$, suggests that the high derivative $(\frac{\partial R_C}{\partial V_G})$ may significantly contribute to the mobility according to Equation (4).

The values of $R_C$ and $R_S$ for the 2P device were derived from the 4PP data using device geometric conversion. $R \cdot W$ of the equivalent $R_C, R_S$ and $R_{tot}$ for the 2P device are plotted in Figure 7d,e,f (labelled as "derived") at comparable channel field strengths (-0.1 V/mm, -1 V/mm, and -20 V/mm). For comparison, the $R_{tot}$ calculated from the 2P output characteristics (labelled "measured") is also shown. The reasonable match between the "derived" and "measured" $R_{tot} \cdot W$ values confirms that the method used to resolve $R_C$ and $R_S$ for the 2P device is reliable.

Reducing the *L/W* ratio significantly decreases $R_S W$ to a similar level as $R_C W$. The large contribution of $R_C$ to $R_{tot}$ leads to an underestimation of FET mobility according to equation (4). This underestimation is more pronounced at low source-drain bias, as shown in Figure 7d, where $R_C W$ can exceed $R_S W$ at $V_G > +50$ V. Increasing the source-drain bias results in smaller $R_C W$ as shown in Figure 7f. Nevertheless, $R_C$ still declines more sharply than $R_S$. Consequently, even at a high source-drain bias, the high derivative $(\frac{\partial R_C}{\partial V_G})$ may remain high and contribute to overestimated $\mu_{sat}$ near the threshold region as shown by Gundlach et al.[9]

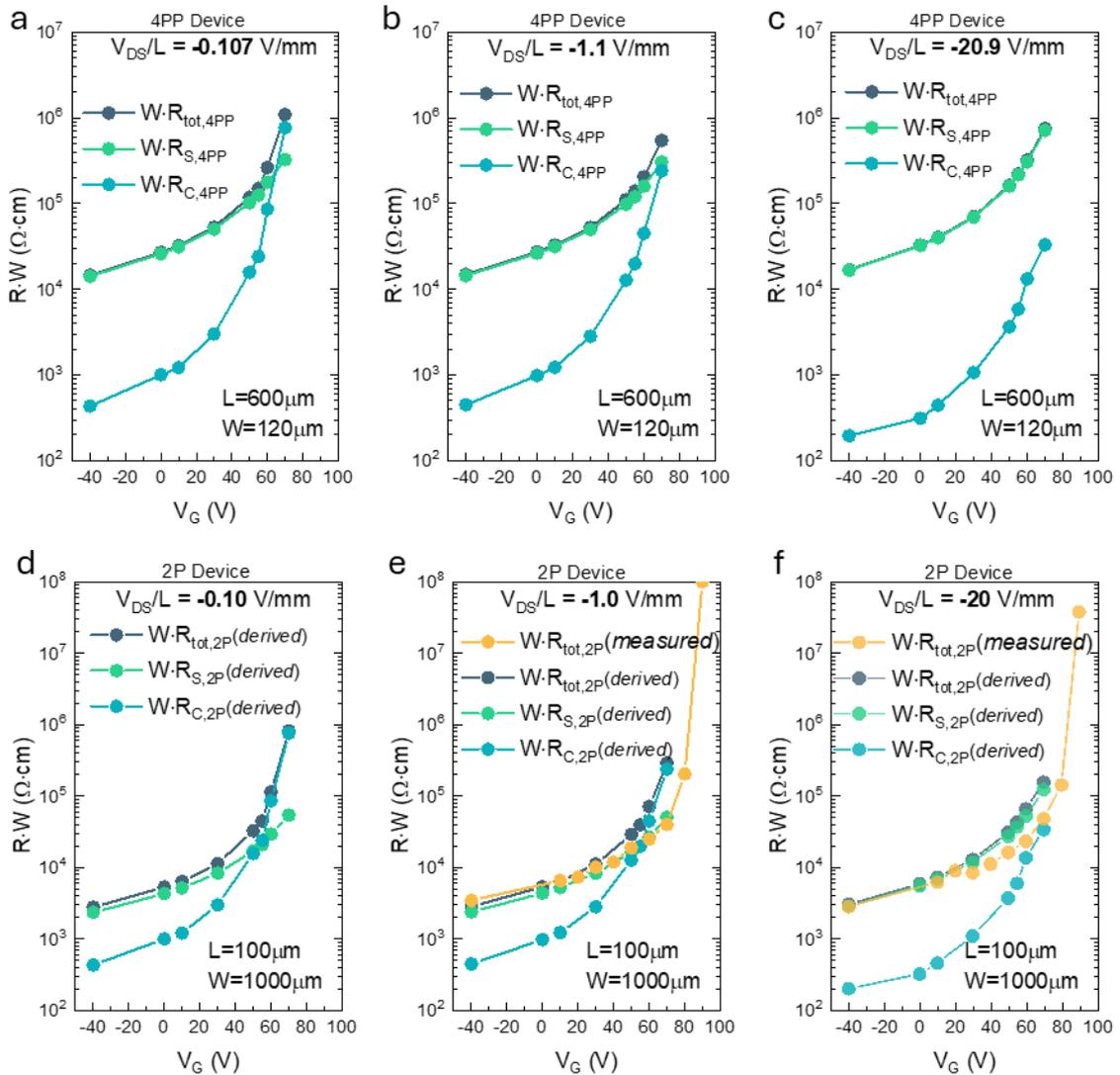

*Figure 7 Gate modulated $R \cdot W$ in 4PP and 2P devices. $R \cdot W$ for $R_C, R_S$ and $R_{tot}$ in 4PP device at different electric field strengths are shown in **(a)** -0.107 V/mm, **(b)** -1.1 V/mm, and **(c)** -20.9 V/mm. Measured and derived resistances for 2P device at **(d)** -0.1 V/mm, **(e)** -1 V/mm, and **(f)** -20 V/mm. The derived $R_S W$ and $R_C W$ for 2P device is calculated through geometric conversion under the assumption that $R_C \cdot W$ and $R_S \cdot W/L$ should remain consistent for both 2P and 4PP devices on the same chip. 2P output characteristics was not measured at $V_{DS}$=-0.01V (-0.10 V/mm).*

These results indicate that $R_C$ is significantly modulated by both $V_{DS}$ and $V_G$. In the linear regime operation, a low $V_{DS}$ can lead to an underestimation of linear FET mobility, while a high $V_{DS}$ can result in a high derivative ($\frac{\partial R_C}{\partial V_G}$), leading to an overestimation of saturation mobility. To achieve reliable mobility extraction in 2P FETs, a large *L/W* ratio is recommended to ensure that $R_S$ is sufficiently greater than $R_C$. However, the best approach to eliminate the influence of contact resistance in mobility extraction is to use the gated-4PP method.

**Temperature Dependence of 4PP and 2P Mobilities**

Measuring the temperature dependence of FET mobility is a key method for determining charge transport mechanisms. We have shown that 2P mobility can vary significantly under different $V_{DS}$ and $V_G$ conditions. In this section, we investigate the impact of $V_{DS}$ on 2P and 4PP mobility at temperatures from 120 K to 295 K. The transfer characteristics of 2P-linear ($V_{DS}$ = -3 V), 2P-saturation ($V_{DS}$ = -60 V), and 4PP ($V_{DS}$= -18 V) measurements are plotted in Figure 8a,b,c. All three conditions display a decrease in $I_{DS}$ at low temperatures, indicating a thermally activated transport mechanism. The extracted mobility at different temperatures for 2P and 4PP devices is plotted in Figure 8d,e with the full datasets of temperature dependent output and mobility shown in Figure S9, Figure S10 and Figure S11.

In the 2P linear regime, mobility increases from approximately 0.1 $cm^2V^{-1}s^{-1}$ (120 K) to approximately 3 $cm^2V^{-1}s^{-1}$ (295 K) and is higher when using larger $V_{DS}$. Saturation mobilities exceed 1 $cm^2V^{-1}s^{-1}$ at all temperatures and also scale up with increasing $V_{DS}$. However, based on earlier resistance analysis, these saturation mobility values are influenced by the large ($\frac{\partial R_C}{\partial V_G}$) in the subthreshold region and may not accurately reflect the temperature dependence of device's intrinsic mobility.

The 4PP mobilities extracted at both high ($V_{DS}$= -18 V) and low $V_{DS}$ ($V_{DS}$= -0.3 V) values overlap consistently across all temperatures. Their temperature dependence shows thermally activated transport near room temperature, which is typically observed in polycrystalline perovskite systems where charge carrier hopping through point defects or grain boundaries is driven by thermal energy.[13,25] Interestingly, below 180 K, the mobility plateaus and even slightly increases at lower temperatures. This trend is less apparent in the 2P results due to more pronounced contact resistance at lower temperatures, but the temperature dependence exhibited by the 4PP mobility is a robust result and not affected by contact resistance.

The temperature dependence of the 4PP mobility is unusual. The near temperature-independent mobility below 180 K cannot be explained by the coexistence of different scattering mechanisms. If there were, for example, a temperature-dependent carrier scattering mechanism, that becomes less effective at high temperature, and a temperature-independent one, this should give rise to a near temperature-independent mobility at high temperatures, not at low temperatures according to Matthiessen's rule. A potential explanation is that the change in the temperature trend of the mobility is related to a phase transition between different tetragonal or orthorhombic structures, which has been reported for the structurally related $FASnI_3$ between near 150 K to 200 K.[29,30] However, a phase transition typically results in an abrupt discontinuity of the mobility or electrical conductivity [6,29,31] and not just a change in the temperature trend. Another possibility is that the observed temperature dependence reflects the temperature dependence of impurity scattering due to charged and/or neutral defects; in traditional band-like semiconductors[32,33] and tin perovskites[34], the mobility due to impurity scattering has been observed to be $\propto T^\gamma$ with $\gamma$ = $^1/_2$ or $^3/_2$ and sometimes approaching a near constant mobility value at low temperatures. A third explanation is that transport involves two parallel conduction pathways, one involving thermally activated transport, for example due to conduction in the valence band being hindered by grain boundaries, and the other one involving temperature-independent transport, for example due to tunnelling within an impurity band. At low temperatures the conduction pathway through the impurity band may become dominant as the temperature activated pathway freezes out. The detailed transport mechanism responsible for the unusual temperature dependence of the mobility will need to be investigated in more detail in the future. It is possible that it is related to other observations reported, such as an irregular PL intensity quenching phenomenon (called negative thermal quenching) in $FASnI_3$ taking place from 185 K to 130 K as reported by Loi et al.[30] and Sun et al.[35]

Figure S12 shows the extracted $R_C \cdot W$, $R_S \cdot W$ and the ratio of $R_C/R_S$ for 4PP and equivalent 2P devices. During cooling, $R_C$ increases from 15%$R_S$ at room temperature to nearly 60% of $R_S$ at 120K for the 2P device. This significant rise in $R_C$ leads to underestimated 2P mobility, which becomes more pronounced at lower temperatures. In contrast, the 4PP results exhibit better consistency, free of contact resistance effects and across the entire temperature range.

We further extracted thermal activation energy ($E_a$) from the temperature dependence of mobility between 295 K and 180 K, as shown in Figure 8f. The 2P saturation mobilities yield a small $E_a$ of about 25 meV at ($V_{DS}$= -60 V), while 2P linear mobility shows a much large $E_a$ of over 110 meV. This large variation in $E_a$ from the same 2P device is due to inaccuracies in mobility extraction caused by contact resistance. Without the influence of contact resistance, 4PP mobility exhibit an $E_a$ of 75 meV.

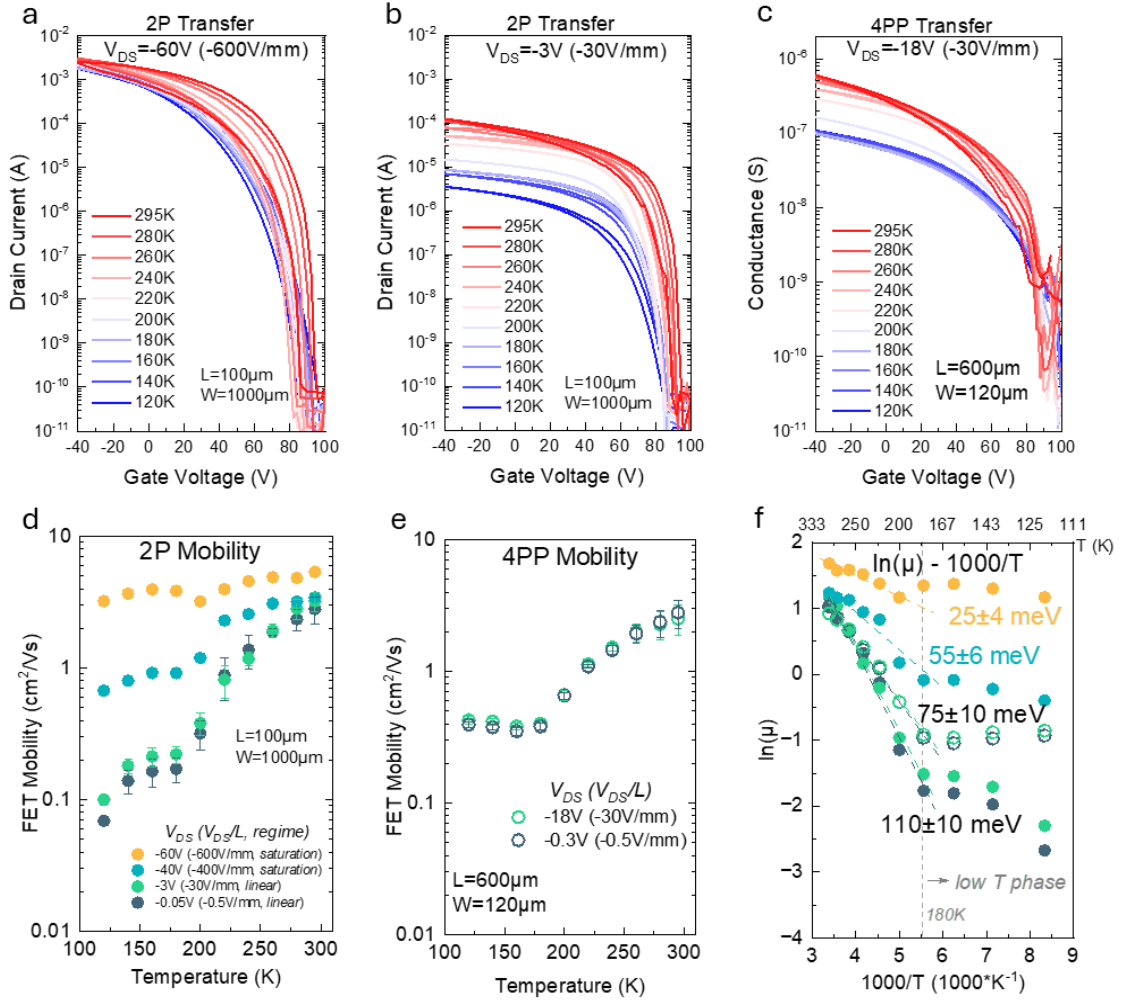

*Figure 8 Temperature Dependent 2P and 4PP FET characteristics*

*Transfer curves of 2P linear regime ($V_{DS}$=-3V) **(a)** and 2P saturation regime ($V_{DS}$=-60V) **(b)** from 120 K to 295 K measured on 2P device (L=100µm, W=1mm). **(c)** Transfer curves from 120 K to 295 K measured on 4PP device at $V_{DS}$=-18V (L=600µm, W=120µm). Extracted 2P **(d)** and 4PP **(e)** FET mobility. 2P linear mobility is averaged at -40V<$V_G$<+50V. 2P-saturation mobility is averaged between $V_G$= $V_{ON}$ + $V_{DS}$ ± 10 V, where $V_{ON}$ is turn on voltage indicated on transfer curve. 4PP mobility is averaged between -40V<$V_G$<+50V. **(f)** Ln(µ) against 1000/T plots for the extraction of activation energy. Dashed line marks 180 K where a phase transition takes place and temperature dependence changes.*

**Discussions and Conclusions**

In this study, we performed gated four-point probe FET measurements on patterned Hall bar devices based on tin perovskite composition $Cs_{0.15}FA_{0.85}SnI_3$ and obtained 4PP FET mobility of 3.3 $cm^2V^{-1}s^{-1}$ independent of $V_{DS}$. We compared these measurements, on the same chip, to gated two-probe measurements of narrow-channel FETs (L/W = 0.1). 2P devices generally exhibit smaller mobility values and a more significant dependence on $V_{DS}$ and $V_G$. This dependence originates from contact resistance ($R_C$). Through 4PP resistance analysis, we resolved $R_C$, $R_S$, and their impact on the total resistance ($R_{tot}$) for both 2P and 4PP devices. Our results indicate that $R_C$ is significantly modulated by both $V_{DS}$ and $V_G$. Specifically, $V_{DS}$ determines injection Schottky barrier thickness, while $V_G$ modulates doping concentration in the contact area. In linear regime operation, a low $V_{DS}$ can lead to an underestimation of linear

FET mobility due to a large voltage drop on $R_C$ if the value of $R_C$ is large, while a high $V_{DS}$ can result in a high derivative ($\frac{\partial R_C}{\partial V_G}$), leading to an overestimation of saturation mobility. When determining the charge transport mechanism through temperature-dependent measurements, we also found that 4PP mobilities are more consistent than 2P linear results as $R_C$ becomes significantly pronounced at low temperatures, and saturation regime 2P mobility here is not reliable as the overestimation seem to be more significant. We found 4PP mobility flattens and slightly increases at temperatures below 180 K, which might suggest that an impurity band conduction pathway dominates at low temperature, when transport pathways that require thermal activation are frozen out. We also found that the FET characteristics in $Cs_{0.15}FA_{0.85}SnI_3$ are affected by ion migration at the dielectric interface, which can be mitigated by operating the device under a pulsed gate or at low temperature.

In conclusion, we have systematically compared the FET characteristics and contact resistance in narrow channel 2P devices and long Hall bar channel 4PP devices based on the tin perovskite composition $Cs_{0.15}FA_{0.85}SnI_3$. Our findings confirm that $R_C$ is highly dependent on both $V_{DS}$ and $V_G$, which can lead to an underestimation of linear 2P mobility at low $V_{DS}$ and an overestimation of saturation 2P mobility at high $V_{DS}$. The most effective approach to eliminate the influence of contact resistance in mobility extraction and determining charge transport mechanisms is to use gated four-point probe (4PP) method.

Our investigation underscores the need for strategies to reduce contact resistance of perovskite FETs, which is crucial for downscaling for circuit integration, and also highlights the importance of measuring contact resistance in order to reliably assess the intrinsic properties of perovskite semiconductors in FETs. Reliable mobility extraction is crucial for better understanding of the charge transport mechanism, that leads, for example, to the observed mobility flattening below 180 K, and for devising strategies to further enhance the carrier mobilities of tin-based perovskite semiconductors.

**Methods**

*Materials*

$SnI_2$ (99.99%, anhydrous, perovskite grade), CsI (99.999%, anhydrous), $SnF_2$ (99%), N,N-Dimethylformamide (DMF, 99.8%, anhydrous), Dimethyl sulfoxide (DMF, 99.9%, anhydrous), Chlorobenzene (99.8%, anhydrous) and PMMA are purchased from Sigma-Aldrich. Formamidinium iodide (99.99%) is purchased from Greatcell Solar Materials.

*Device substrate fabrication*

$SiO_2$ wafers (300nm dry thermal $SiO_2$ on 525μm Silicon) are pre-cut and patterned with 4PP and 2P FET patterns by photolithography method. 4 nm titanium and 22nm gold are deposited by thermal evaporation in high vacuum ($2 \cdot 10^{-6}$ mbar). Substrates are lifted-off in N-Methyl-2-pyrrolidone (NMP), cleaned by isopropanol and oxygen plasma (300 watt, 10 mins).

*Perovskite deposition and patterning*

Deposition method of tin perovskite $Cs_{0.15}FA_{0.85}SnI_3$ is described below. 0.3 mmol $SnI_2$ and 0.3×12.5% mmol $SnF_2$ are dissolved in 400 μL DMF-DMSO cosolvent (3:1) and stirred at room temperature for 30 minutes (forming 0.75 M solution). The solution is then transferred into a vial containing 15%×0.3×105% mmol CsI and 85%×0.3×105% FAI powders. The solution is further stirred for 30 minutes and filtered through a PTFE membrane filter (pore size 0.2μm). 20 μL solution was spin coated onto a substrate at 5000 rpm for 40 seconds, with 100 μL Chlorobenzene drop cast at the first 9th second. The film is then annealed at 100°C for 10 minutes. For encapsulation, 20 μL PMMA (dissolved in Chlorobenzene at 60 mg/ml) is then spin coated onto the perovskite film at 2000 rpm for 30 seconds and annealed at 100°C for 10 minutes. All the above operations are handled in a nitrogen glovebox with oxygen and water level lower than 0.5ppm. The films are initially rough-patterned by cocktail sticks in a glovebox and then fine-patterned by metallic tips in a vacuum probe station.

*Materials characterisations*

SEM imaging was carried out in high vacuum ($<4 \times 10^{-6}$ mbar) by a LEO GEMINI 1530VP FEG-SEM using 20 kV acceleration voltage. XRD measurements were collected using a Bruker D8 Advance powder X-ray diffractometer with Cu Kα (λ = 1.54 Å) radiation. For photophysical measurements, perovskite films are deposited on glass substrate and encapsulated by a layer of PMMA and a capping glass glued by epoxy resin. Steady-state UV-Vis Absorption spectra were measured using an Agilent Cary 7000 UV-vis-NIR spectrophotometer. Steady-state photoluminescence spectra were measured on an Edinburgh Instruments FLS1000 fluorimeter with a 450 W continuous xenon arc lamp at 405 nm excitation. The data was acquired separately in the visible and near-infrared ranges and merged. Time-correlated single photon counting (TCSPC) plots were measured on an Edinburgh Instruments FLS1000 fluorimeter. A picosecond pulsed diode laser at 404 nm (HPL-405) was used (repetition rate of 5 MHz), which was connected using a coupling flange.

*Electrical measurements*

2P and 4PP devices were on a same chip. Device chips was loaded in a Lakeshore TTP4 cryogenic probe station and pumped to high vacuum ($<1 \times 10^{-5}$ mbar). Electrical measurements were performed by an Agilent 4155C Semiconductor Parameter Analyzer with an experimental control platform FetCh (https://github.com/OE-FET/FetCh) based on experimental instrumentation library JISA[36]. 2P and 4PP transfer characteristics were measured with a gate bias voltage step of 2V and a current integration time of 20 ms. 4PP pulsed gate bias measurements were performed with a current integration time of 20ms and pulse interval time of 0.5 s. 2P and 4PP output characteristics were measured with continuous gate and source-drain biases in a current integration time of 20 ms. For temperature dependent measurements, device was stabilized for 10 minutes at each temperature point. Temperature-dependent measurements of 4PP and 2P devices are separately measured in different temperature cycles. Measurements are carried out in complete dark and undisturbed during the whole temperature cycle.

**Acknowledgements**

We acknowledge funding from the Engineering and Physical Sciences Research Council for a programme grant (EP/W017091/1) and the European Research Council (ERC) for an Advanced Grant (101020872). Y.Z. acknowledges the support from Cambridge University Postgraduate Hardship Funding, Cambridge University PGR Covid-19 Assistance Scheme, UKRI Covid-19 Phase II Doctoral Extension Funding, Darwin College Hardship Funding, the EPSRC Centre for Doctoral Training in Graphene Technology and Cambridge Philosophical Society for the research studentship. X.W.C. thanks the Agency for Science, Technology and Research (Singapore) for the National Science Scholarship. C.Z. thanks the Winton Programme for Physics of Sustainability for funding support. K.D. acknowledges the support of the Cambridge Trust for the Cambridge India Ramanujan Scholarship and Cambridge Philosophical Society for the research studentship. S.P.S. acknowledges funding support from DAE, Government of India through project RIN 4001. H.S. gratefully acknowledges the Royal Society for a Royal Society Research Professorship.

**Conflict of Interest**

The authors declare no conflict of interest.

**Author contributions**

Y.Z. and S.P. conceptualized the idea. Y.Z. designed the experiments, fabricated the devices, performed electrical measurements, XRD, SEM, and wrote the manuscript with inputs from S.P., X.R. and H.S. X.W.C. performed the photophysical measurements. S.P., C.Z., R.M., S.P.S. and K.D. fabricated devices in the preliminary stage of the project. H.S. supervised the project. All authors discussed the work and revised the manuscript.

**Supplementary Information**

**for**

**Critical assessment of contact resistance and mobility in tin perovskite semiconductors**

Youcheng Zhang, Stefano Pecorario, Xian Wei Chua, Xinglong Ren, Cong Zhao, Rozana Mazlumian, Satyaprasad P. Senanayak, Krishanu Dey, Sam Stranks and Henning Sirringhaus[1]*

**Supplementary Section 1**: FET pattern designs

**Supplementary Section 2**: Transfer Characteristics of 4PP devices

**Supplementary Section 3**: 2P vs. 4PP devices at 220K

**Supplementary Section 4**: 4PP channel pinch-off at larger source-drain bias

**Supplementary Section 5**: Reproduction of 2P and 4PP results on a different chip

**Supplementary Section 6**: Temperature dependent 4PP output characteristics

**Supplementary Section 7**: Temperature dependence of 4PP and 2P mobilities

**Supplementary Section 8**: Temperature dependence of contact and sheet resistances

**Supplementary Section 1**: FET pattern designs

Detailed geometry of the 4PP and 2P FET devices are shown Figure S 1. The 4PP device is approximately 2.2mm long and 1.8mm wide (Figure S1a). The centered Hall bar channel is 0.6mm long and 0.12mm wide (Figure S1b). 8 voltage probes are placed on both sides of the Hall bar. The probes have a width of 5μm. For the 4PP measurements in this study, voltage probes near the source and drain electrodes are used as voltage probes with voltage probe distance $D_{4P}$=0.57mm. In Figure S1c, the 2P FET devices have interdigitated electrodes with $L$=100μm and W=$1$mm.

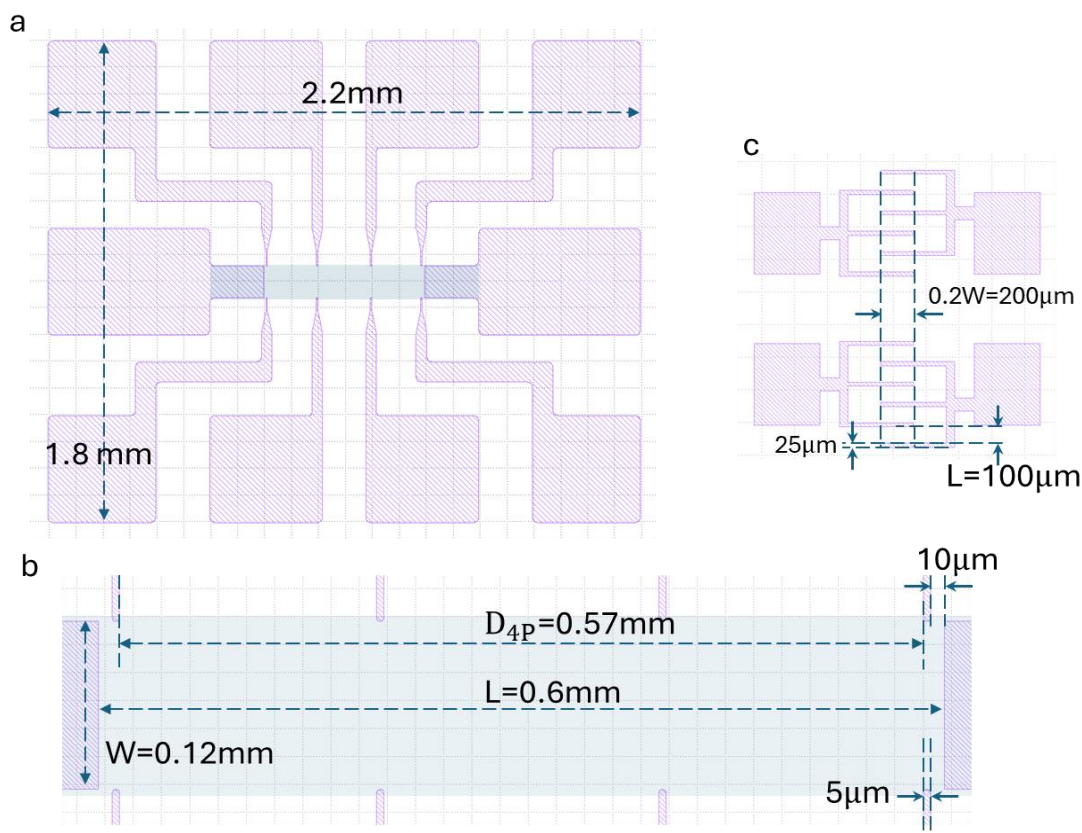

*Figure S 1 FET pattern designs. (a) Four-point-probe (4PP) FET contact pattern design. Hall bar area is marked in grey. (b) View of the hall bar channel area with geometry parameters annotated. (c) Two-probe (2P) FET contact pattern design and geometry.*

**Supplementary Section 2: Transfer Characteristics of 4PP devices**

Transfer Characteristics of the 4PP device in corresponding to the conductance plots shown in Figure 2 and Figure 3 are replotted to $I_{DS}$ in Figure S2. For the transfer curves measured with continuous gate bias at 295K (Figure S2a,d), it is clear that $I_{DS}$ is highly modulated by $V_{DS}$ that the ON current varies from 0.3 μA to 16 μA as $V_{DS}$ increases from -0.3 V to -18 V. Gate leakage current remains below 1nA. Turn-on voltage $V_{ON}$ shifts from +85V to near +100V. The transfer curves are slightly superlinear when the current is high. Hysteresis loop direction changes from counterclockwise to clockwise at $V_G$<0V. Measuring with a pulsed gate bias at the same temperature (Figure S2b,e) results in slightly smaller $I_{DS}$ of 10µA and smaller $V_{ON}$ of +85V for all $V_{DS}$ values. The curves display better linearity and hysteresis is counterclockwise, suggesting ion migration at the dielectric interface is mitigated. Transfer characteristics measured at 220K with a continuous gate bias show negligible hysteresis and textbook-like linearity, suggesting complete suppression of ion migration at low temperature.

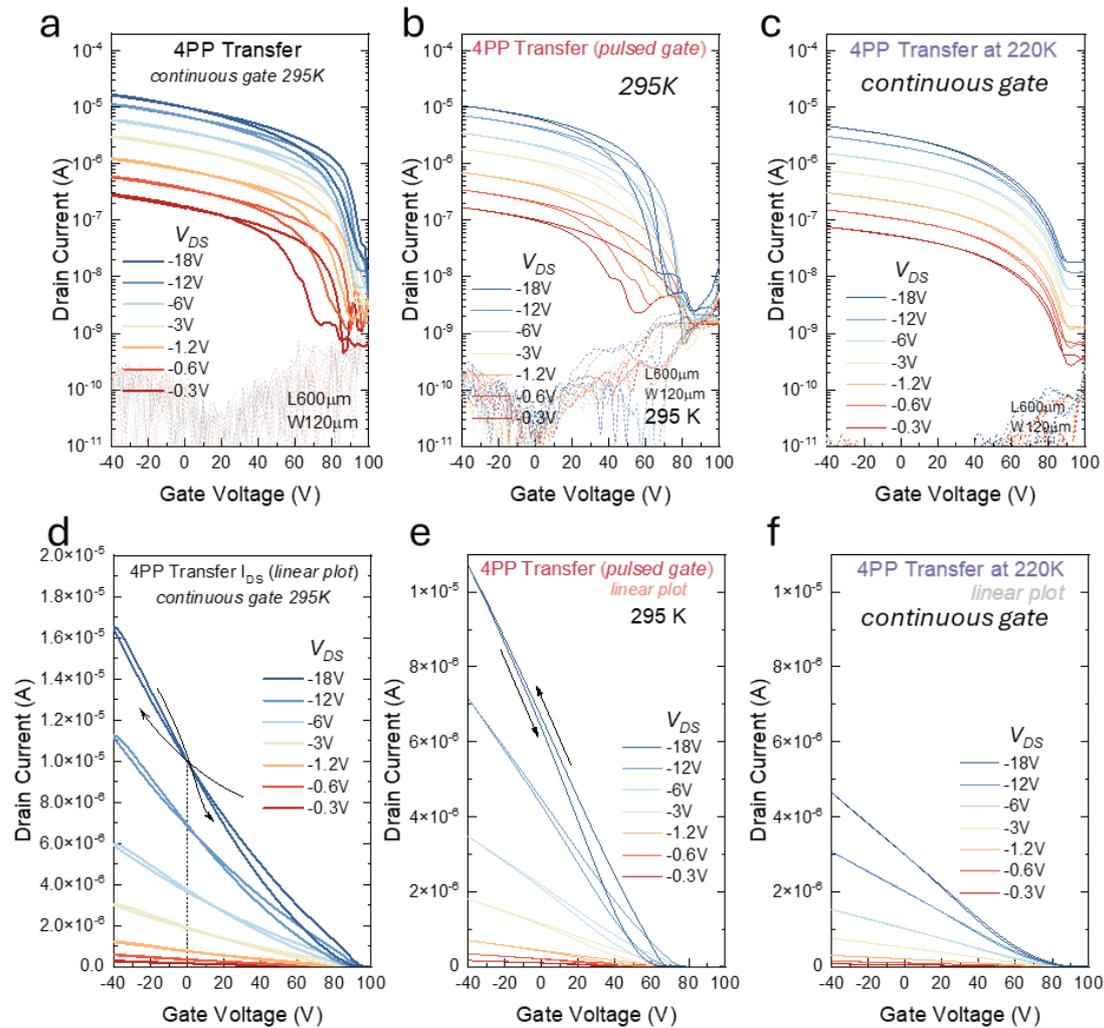

*Figure S 2 FET Characteristics from the gated four-point-probe (4PP) devices plotted in $I_{DS}$ instead of conductance.* **(a)** Transfer characteristics measured in continuous mode at 295K, corresponding to data shown in Figure 2. **(b)** Transfer characteristics measured in pulsed mode at 295K, corresponding to data shown in Figure 3. **(c)** Transfer characteristics measured in continuous mode at 220K, corresponding to data shown in Figure 3. Linear scale plots of current are shown for **(d)** continuous mode at 295K, **(e)** pulsed mode at 295K and **(f)** continuous mode at 220K. Gate leakage current at different scans is shown in faint dash line. Hysteresis loop direction is marked by the black arrows. *L=600 µm, W=120 µm, $D_{4P}$=570 µm.*

Change of voltage values recorded on voltage probe 1 and 2 ($V_{P1}$ and $V_{P2}$) during 4PP transfer characteristics scans are plotted in Figure S3. Voltage probe 1 near the drain electrode displays voltage values nearly equal to $V_{DS}$, while voltage probe 2 near the source electrode displays a significant change at $V_G$ near the threshold region. T*his indicates that contact resistances are asymmetrical at source and drain sides, that source side resistance is much larger than drain side resistance near the threshold region.* A sharp jump of $V_{P2}$ near $V_G$=+100V is due to a pronounced gate leakage.

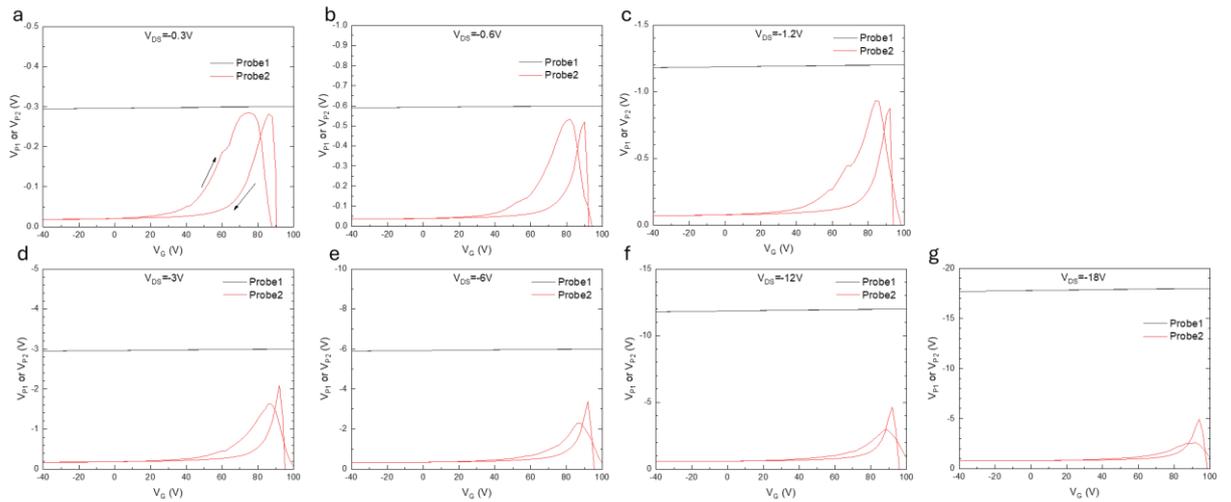

*Figure S 3 Change of $V_{P1}$ and $V_{P2}$ modulated by gate voltage in transfer scans measured at different $V_{DS}$. L=600 μm, W=120 μm, $D_{4P}$=570 μm.*

**Supplementary Section 3: 2P vs. 4PP devices at 220K**

Transfer characteristics of 2P devices at 220K are shown in Figure S4. The transfer curves show negligible hysteresis and lower $I_{DS}$. Double slope feature is found on $|V_{DS}| \leq 10V$, which is an indication of pronounced contact resistance effect. Correspondingly, the extracted linear FET mobilities exhibit a hump near $V_G$=+50V, which should be resulted from a high $\frac{\partial R_C}{\partial V_G}$ term at low temperature as suggested by Equation 4 in the main paper. Applying a high $V_{DS}$ of -20V helps mitigate the double slope effect and reduce the mobility hump, but the characteristics cannot be treated as linear regime at such a high $V_{DS}$. 2P saturation mobility also scales with $V_{DS}$ that at $V_{DS}$=-60V the peak saturation FET mobility could reach at 4.5 $cm^2V^{-1}s^{-1}$. Output curves are superlinear at $V_G$=-40V, suggesting charge injection is limited by a Schottky barrier.[1]

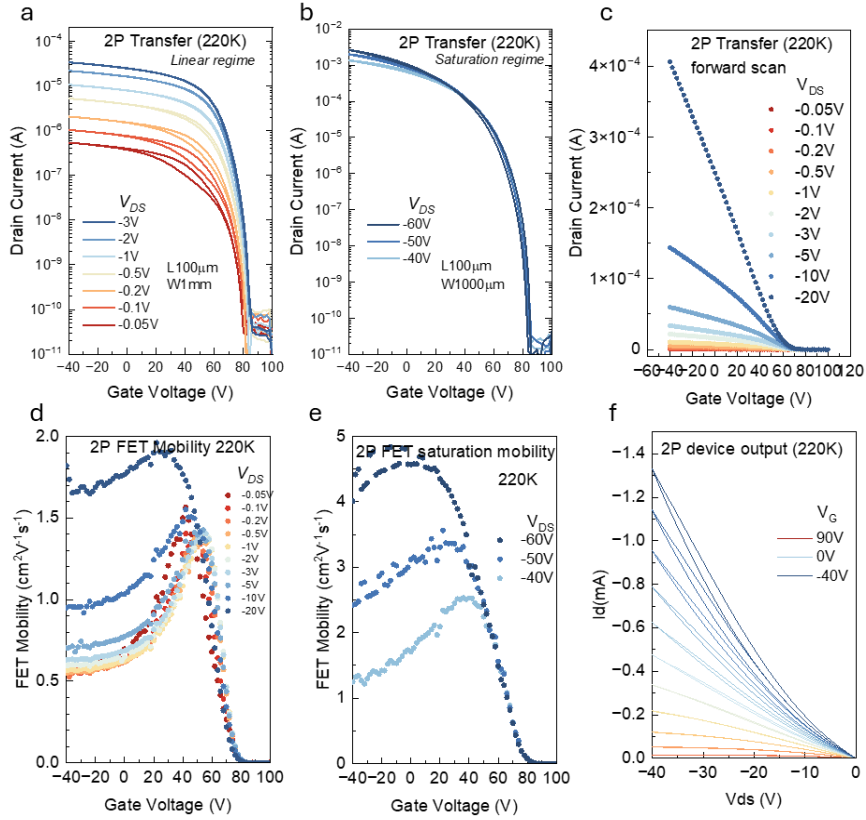

*Figure S 4 2P FET characteristics measured at 220K.* **(a)** Linear regime and **(b)** saturation regime transfer characteristics. **(c)** Linear regime transfer characteristics with $I_{DS}$ plotted in linear scale. **(d)** Linear regime and **(e)** saturation regime FET mobility versus gate voltage. **(f)** output characteristics. $L$=100 $\mu$m. $W$=1mm.

We further extract 2P saturation mobility by fitting the $\sqrt{I_{DS}}$ curves as shown in Figure S5a,b. The regions been fitted are near the saturation regime criteria at $V_{DS}+V_{TH}<V_G<V_{TH}$.[2] We fitted $\sqrt{I_{DS}}$ data points between +20V< $V_G$ <+65V for curves measured at 295K and +10V< $V_G$ <+55V for curves measured at 220K. This method yields more accurate saturation mobility values with reasonable error ranges.

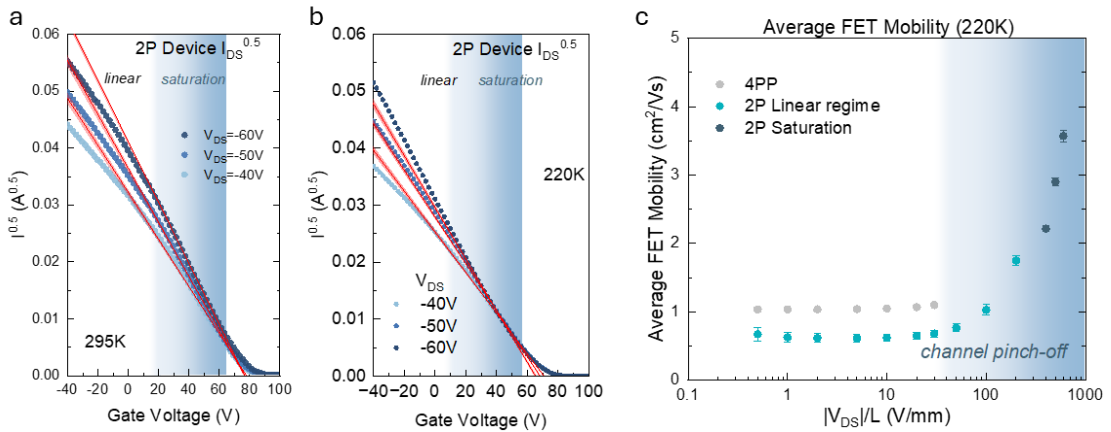

*Figure S 5 Linear fitting of $\sqrt{I_{DS}}$ for 2P device saturation mobility extraction at **(a)** 295K and **(b)** 220K. Saturation regime is marked by blue colour on the graphs. **(c)** Comparison of 2P and 4PP device mobility values at 220K for different channel electric field strength $|V_{DS}|/L$.*

Figure S5c compares the 4PP, 2P linear and 2P saturation mobility values at 220K. 4PP mobility remains at 1 cm$^2$V$^{-1}$s$^{-1}$ and is independent of $|V_{DS}|/L$. 2P linear mobility remains stable at 0.6 cm$^2$V$^{-1}$s$^{-1}$ until the onset of channel pinch-off. Extracting mobility by the linear method for 100V/mm and 200V/mm results in overestimated mobility values as high as 1.8 cm$^2$V$^{-1}$s$^{-1}$. Devices measured at these electric field strengths cannot be treated as pure linear regime or pure saturation regime. 2P saturation mobility is also higher at high $|V_{DS}|/L$. Overall, the extracted mobility values obtained at 220K show smaller error bars, suggesting ion migration related non-idealities is effectively mitigated at low temperature.

**Supplementary Section 4**: 4PP channel pinch-off at larger source-drain bias

4PP device is not measured at electric field strength higher than 30V/mm due to channel pinch off, that voltage does not drop linearly across the channel.[3] Figure S6 shows the voltage values measured at the two voltage probes when performing the transfer characteristics at $V_{DS}$=-20V and $V_{DS}$=-40V. In Figure S6a, $V_{P1}$ nearly equals to -20V at $V_{DS}$=-20V but is limited to -23V at $V_{DS}$=-40V. The latter is an indication of channel pinch off. Therefore, it is not possible to extract 4PP mobility from 4PP transfer measurements carried out at $|V_{DS}|/L$ over 30V/mm.

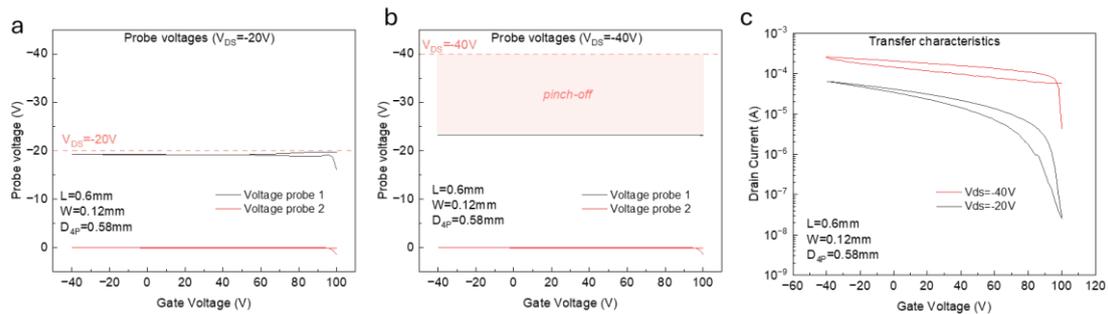

*Figure S 6 4PP channel pinch-off at larger source-drain voltage. Voltage values recorded on two FPP voltage probes in the transfer curve scans at (a) V$_{DS}$=-20V (-33.3 V/mm) and (b) V$_{DS}$=-40V (-66.7 V/mm), where Vp1 saturates at -23.3 V due to channel pinch-off. (c) Transfer characteristics of the two scans. ON/OFF ratio was largely reduced at V$_{DS}$=-40V. L=600 μm, W=120 μm, D$_{4P}$=570 μm.*

**Supplementary Section 5:** Reproduction of 2P and 4PP results on a different chip

The same set of measurements were performed on a different chip to verify the results. In Figure S7a,b,c, the 4PP transfer characteristics exhibit highest ON current of 12μA, slightly superlinear transfer curves, and 4PP mobility of 3±1.5 cm$^2$V$^{-1}$s$^{-1}$ independent of $V_{DS}$. In Figure S7f, the extracted 2P linear mobility is dependent on both $V_{DS}$ and $V_G$. 2P device also exhibits a high saturation ON current of 5mA, but the device is not turned off in the reverse scan in Figure S7g. The forward scan yields a peak mobility of 8.5±1 cm$^2$V$^{-1}$s$^{-1}$ in Figure S7h. Comparing 4PP mobility against 2P linear and saturation mobilities, the 4PP mobility is independent on channel electric field strength, while the 2P mobilities are highly $V_{DS}$-dependent at $\frac{|V_{DS}|}{L}$ >1V/mm. 4PP mobility maintains at 3.2±1 cm$^2$V$^{-1}$s$^{-1}$, which is higher than 2P linear mobility at $\frac{|V_{DS}|}{L}$ ≤30V/mm. These results highly resemble the FET results shown in the main paper.

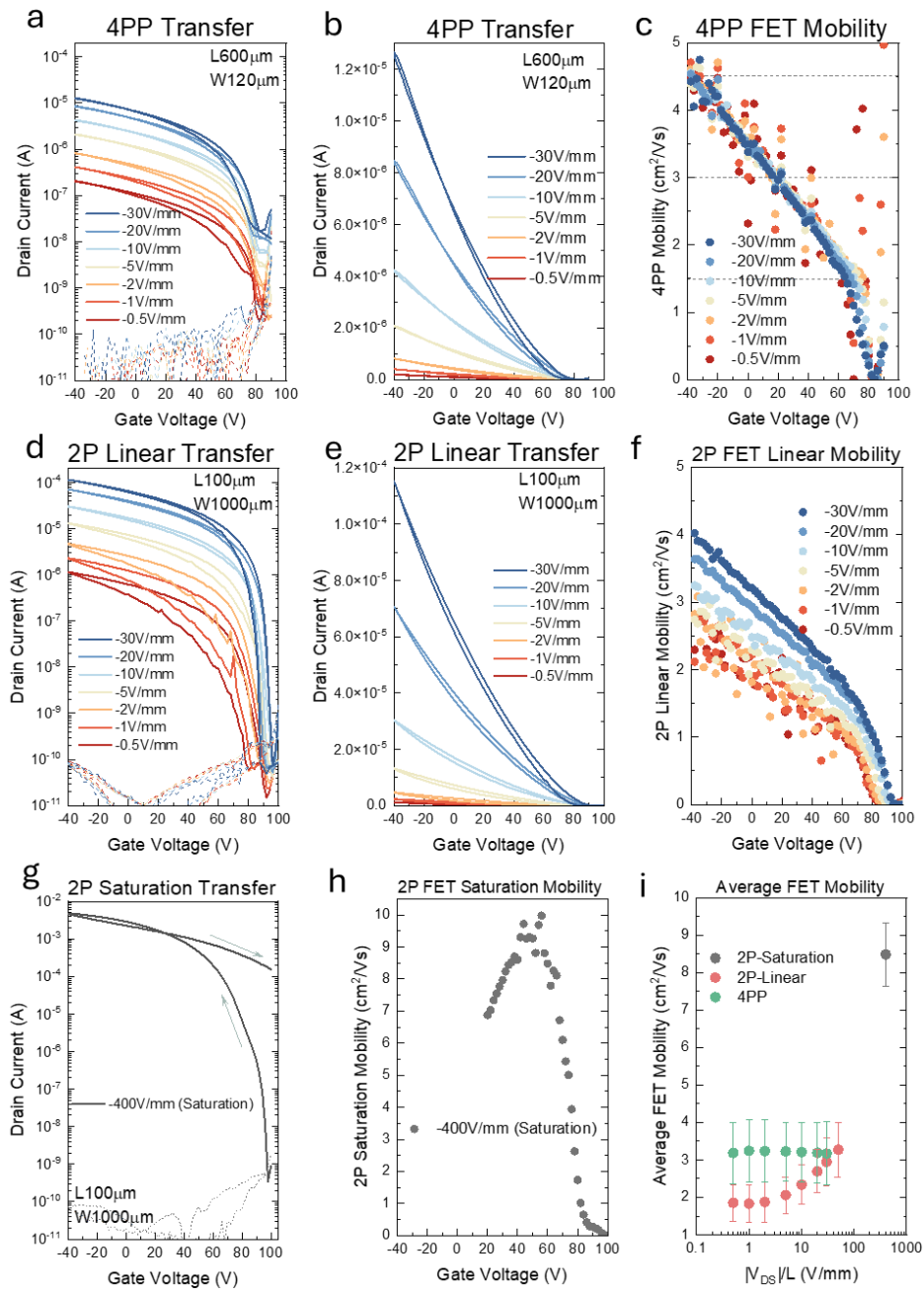

Figure S 7 Repeat of 4PP and 2P FET measurements on a different chip. Chip is fabricated from a different batch. Perovskite composition is unchanged. 4PP device transfer characteristics plots in **(a)** logarithm and **(b)** linear scales. **(c)** extracted 4PP FET mobility at different $V_{DS}$ values. 2P device linear regime transfer characteristics plots in **(d)** logarithm and **(e)** linear scales and **(f)** extracted 2P linear mobility values. **(g)** 2P device saturation transfer characteristics and **(h)** extracted saturation mobility values. **(i)** Comparison of 2P and 4PP mobility extracted at different FET channel electric field strengths.

**Supplementary Section 6:** 4PP output characteristics for contact resistance extraction

Channel sheet resistance ($R_S$) and contact resistance ($R_C$) can be extracted from 4PP output characteristics. We aim to investigate the modulation effect of $V_G$ and $V_{DS}$ on $R_S$ and $R_C$. We also want to correlate the modulation with channel electric field strength $\frac{|V_{DS}|}{L}$. Therefore $V_{DS}$ is applied in equal exponential increments. The output characteristics are shown in Figure S8a,b and voltage measured by the two voltage probes are shown in Figure S8c,d,e,f. $V_{P1}$ looks equal to $|V_{DS}|$, whereas $V_{P2}$ shows a more pronounced modulation by both $V_G$ and $V_{DS}$.

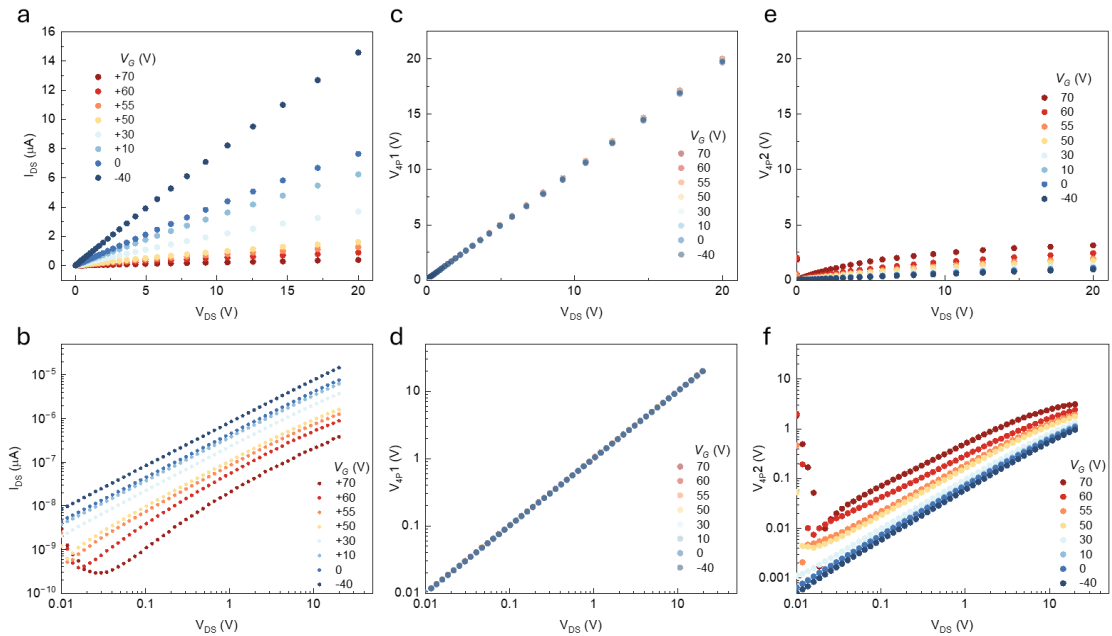

Figure S 8 4PP device Output characteristics for the extraction of $R_S$ and $R_C$. Output characteristics plotted in (a) linear and (b) logarithmic scales. Change of voltage probe 1 $V_{4P1}$ plotted in (b) linear and (d) logarithmic scales. Change of voltage probe 2 $V_{4P2}$ plotted in (e) linear and (f) logarithmic scales. Measurement is done at 295K. L=600 μm, W=120 μm, $D_{4P}$=570 μm.

**Supplementary Section 7**: Temperature dependence of 4PP and 2P mobilities

4PP mobility is extracted for transfer curves measured at $V_{DS}$ =-3V and -18V over a temperature range from 295K to 120K in Figure S9a,b. The mobility dependence on $V_G$ is effectively absent at low temperatures. During cooling, mobility drops from 3 $cm^2V^{-1}s^{-1}$ at 295K to 0.4 $cm^2V^{-1}s^{-1}$ at 180K, and stabilizes around 0.4 $cm^2V^{-1}s^{-1}$ upon further cooling.

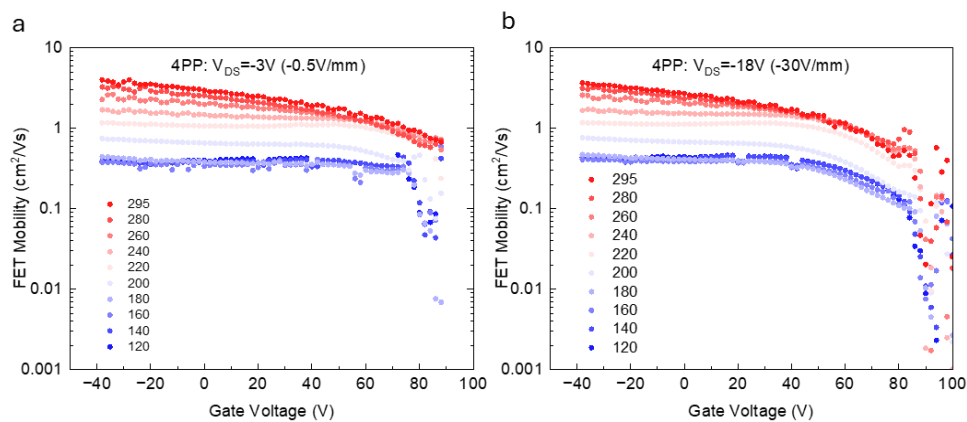

Figure S 9 Temperature and gate-voltage dependence of 4PP mobilities extracted from transfer characteristics measured at (a) $V_{DS}$=-3V and (b) $V_{DS}$=-18V.

2P linear and saturation mobility values are also extracted for transfer curves measured at $V_{DS}$=-0.05V, -3V, -40V and -60V from 295K to 120K in Figure 10. Linear mobility also drops from approximately 3 $cm^2V^{-1}s^{-1}$ at 295K to below 0.2 $cm^2V^{-1}s^{-1}$ at 120K. The extracted 2P saturation mobility against gate voltage graph looks more chaotic. Instead, we chose to extract these mobilities by fitting their $\sqrt{I_{DS}}$ curves in Figure S10e,f, only in the saturation regime. These values are plotted in Figure 8d in the main paper.

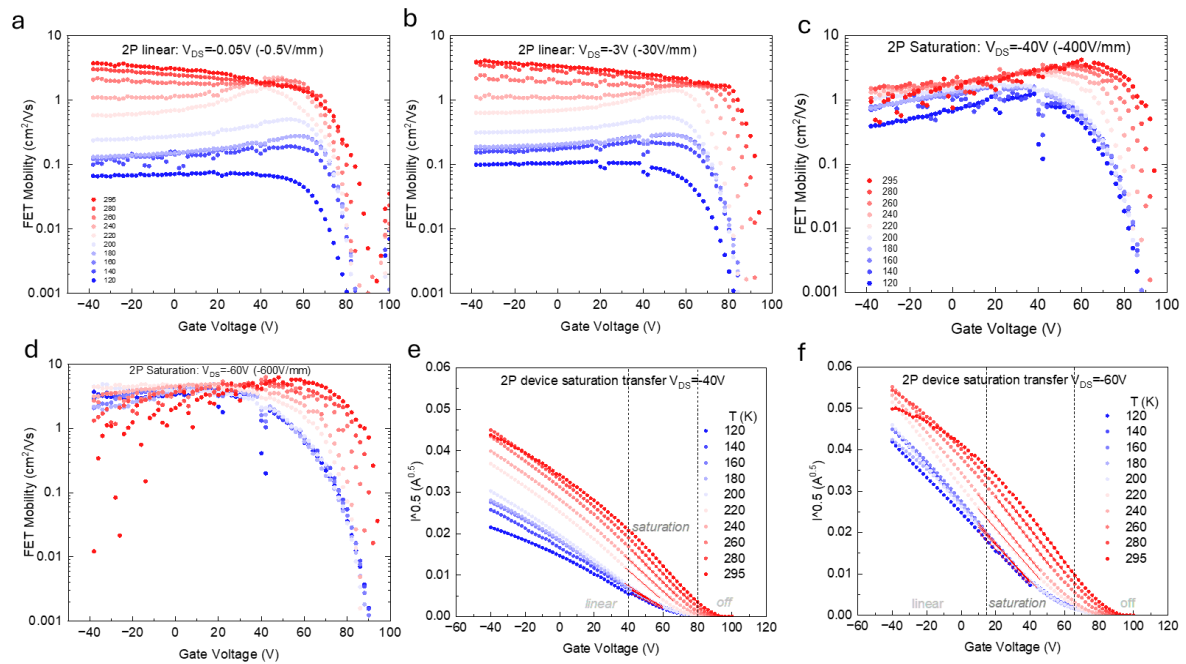

*Figure S 10 Temperature and gate-voltage dependence of 2P mobilities extracted from transfer characteristics measured at linear regime (a) $V_{DS}$=-0.05V, (b) $V_{DS}$=-3V, and saturation regime (c) $V_{DS}$=-40V, (d) $V_{DS}$=-60V. Plots of $\sqrt{I_{DS}}$ for (e) $V_{DS}$=-40V and (f) $V_{DS}$=-60V. Onsets of saturation regime are marked by gray dashed lines.*

Temperature dependence of 2P and 4PP mobility obtained at the same $V_{DS}/L$ is compared for -30V/mm and -0.5V/mm in Figure S11a,b. For both $V_{DS}/L$, it is clear that 4PP mobility stabilizes below 180K, while 2P mobility keeps decreasing.

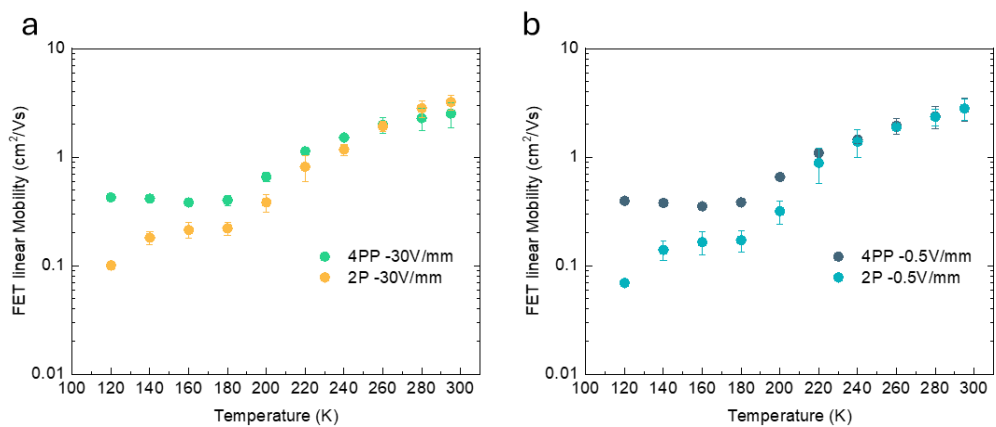

*Figure S 11 Comparison of temperature dependence of 4PP and 2P FET linear mobility extracted at the same electric field strength of (a) -30V/mm and (b) -0.5V/mm.*

**Supplementary Section 8**: Temperature dependence of contact and sheet resistances

Temperature dependence of $R_S \cdot W$ and $R_C \cdot W$ are extracted at $V_G$=0V and $V_{DS}$=-1V to -10V for the 4PP device. $R_S \cdot W$ increases with temperature until 180K. $R_C \cdot W$ first drops from 295K to 260K, then increases until 160K and stabilizes. $R_S \cdot W$ is over 10 times higher than $R_C \cdot W$ for the long Hall bar 4PP device here. Therefore, $R_C/R_S$ is less than 10% for all temperature ranges, as shown in Figure S12b. If converted $R_C/R_S$ to the geometry of 2P FETs with $L$=100µm and $W$=1mm, $R_C$ could take up 60% of $R_S$ at low temperatures. This suggests 2P FETs with short channel lengths is more susceptible to contact resistance. Increasing the channel length could mitigate $R_C$. Figure S12c demonstrates the charge carrier flow pathway from source to the drain through processes 1., 2., and 3. Charge carriers are transported at the dielectric interface, and are injected and extracted through the source and drain contacts. As suggested by processes 1 and 3, contact resistance at each side is the series sum of two terms: a interfacial resistance associated with charge carrier crossing the injection barrier through the metal electrode edge, and a semiconductor bulk resistance associated with charge carriers travelling between metal edge to the dielectric interface.[4] Therefore, $R_C$ is also affected by the properties of the semiconductor bulk. In our case here, $R_C$ shows a similar temperature dependence to $R_S$ that they are both featured by a plateau below 180 K as shown in Figure S12a.

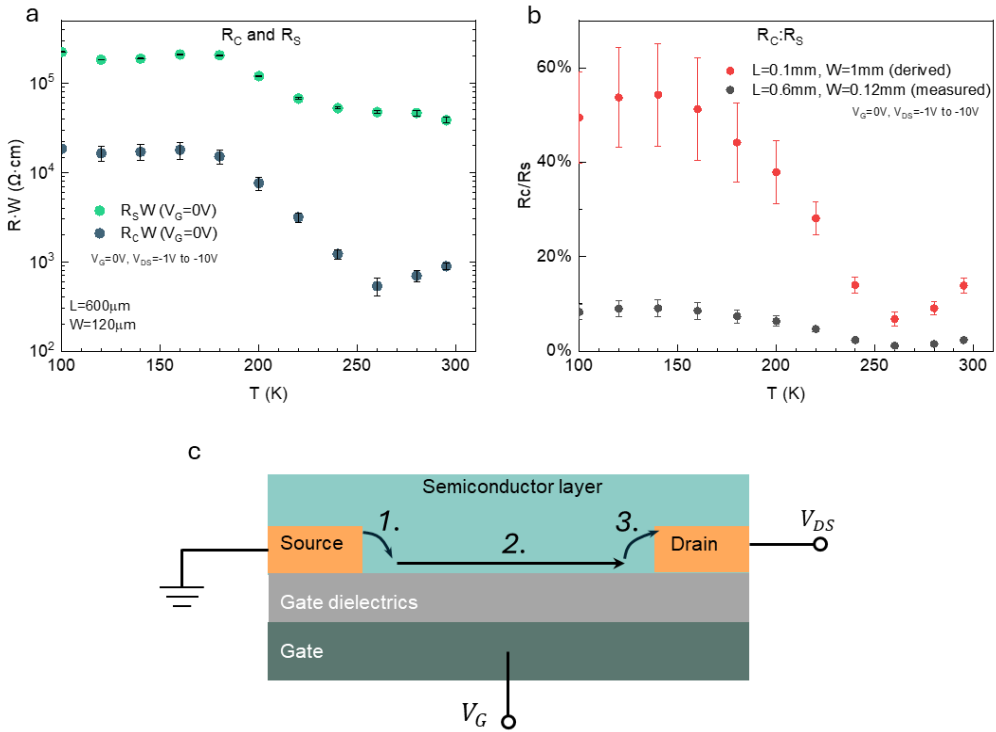

*Figure S 12 Temperature dependence of (a) $R_S \cdot W$ and $R_C \cdot W$ and (b) the ratio of $R_C/R_S$ for 2P and 4PP devices. $R_S \cdot W$ and $R_C \cdot W$ are extracted from output characteristics measured at each temperature with $V_{DS}$ sweeps from 0V to -20V. RW values are averaged from $V_{DS}$=-1V to -10V at $V_G$=0V. For 2P device, geometric conversion is applied to calculated $R_{C,2P}W_{2P}$ and $R_{S,2P}W_{2P}$ by the formula $R_{C,2P}W_{2P}=R_{C,4PP}W_{4PP}$ and $R_{S,2P}W_{2P}=(100\ \mu m/600\ \mu m) \times R_{S,4PP}W_{2P}$. This is done under the assumption that $R_C \cdot W$ and $R_S \cdot W/L$ should remain consistent for both 2P and 4PP devices on the same chip. (c) Schematics for illustrating the origin of contact resistance.*